\documentclass[twocolumn,aps,prl,superscriptaddress]{revtex4-1}

\usepackage{amsmath}
\usepackage{amssymb}
\usepackage{graphicx}
\usepackage{upgreek}
\usepackage{xfrac}
\usepackage{soul}
\usepackage{bm}
\usepackage{xcolor}
\definecolor{orange}{RGB}{235, 129, 0}
\usepackage{comment}
\usepackage{enumitem}

\newlist{UR}{enumerate}{1}
\setlist[UR]{label=M\arabic*.}

\usepackage[bookmarks=true,
            bookmarksnumbered=false,
            bookmarksopen=true,
            breaklinks=true,
            pdfborder={0 0 0},
            final=true,
            backref=none,
            colorlinks=true,
            linkcolor=blue,
            menucolor=blue,
            citecolor=blue,
            urlcolor=blue]{hyperref}

\renewcommand{\mu}{\upmu}
\newcommand{\um}{\mu\textrm{m}}			
\newcommand{\iums}{\mu\textrm{m}^{-2}} 	
\newcommand{\LGOO}{LG_{00}}				

\makeindex

\begin{document}

\title{Highly nonlinear trion-polaritons in a monolayer semiconductor} 

\author{R. P. A. Emmanuele}
\affiliation{Department of Physics and Astronomy, The University of Sheffield, Sheffield, S3~7RH, United Kingdom}

\author{M. Sich}
\affiliation{Department of Physics and Astronomy, The University of Sheffield, Sheffield, S3~7RH, United Kingdom}

\author{O. Kyriienko}
\email{kyriienko@ukr.net}
\affiliation{Department of Physics and Astronomy, University of Exeter, Stocker Road, Exeter EX4 4QL, UK}
\affiliation{Department of Nanophotonics and Metamaterials, ITMO University, St. Petersburg, 197101, Russia}

\author{V. Shahnazaryan}
\affiliation{Department of Nanophotonics and Metamaterials, ITMO University, St. Petersburg, 197101, Russia}

\author{F. Withers}
\affiliation{College of Engineering, Mathematics and Physical Sciences, University of Exeter, Exeter, EX4 4QF, United Kingdom}

\author{A. Catanzaro}
\affiliation{Department of Physics and Astronomy, The University of Sheffield, Sheffield, S3~7RH, United Kingdom}

\author{P.~M. Walker}
\affiliation{Department of Physics and Astronomy, The University of Sheffield, Sheffield, S3~7RH, United Kingdom}

\author{F.~A. Benimetskiy}
\affiliation{Department of Nanophotonics and Metamaterials, ITMO University, St. Petersburg, 197101, Russia}

\author{M.~S. Skolnick}
\affiliation{Department of Physics and Astronomy, The University of Sheffield, Sheffield, S3~7RH, United Kingdom}
\affiliation{Department of Nanophotonics and Metamaterials, ITMO University, St. Petersburg, 197101, Russia}

\author{A.~I. Tartakovskii}
\affiliation{Department of Physics and Astronomy, The University of Sheffield, Sheffield, S3~7RH, United Kingdom}

\author{I.~A. Shelykh}
\affiliation{Science Institute, University of Iceland, Dunhagi-3, IS-107 Reykjavik, Iceland}
\affiliation{Department of Nanophotonics and Metamaterials, ITMO University, St. Petersburg, 197101, Russia}

\author{D.~N. Krizhanovskii}
\affiliation{Department of Physics and Astronomy, The University of Sheffield, Sheffield, S3~7RH, United Kingdom}
\affiliation{Department of Nanophotonics and Metamaterials, ITMO University, St. Petersburg, 197101, Russia}
\email{d.krizhanovskii@sheffield.ac.uk}


\maketitle

\section{\label{sec:intro}}
\textbf{Highly nonlinear optical materials with strong effective photon-photon interactions (Kerr-like nonlinearity) are required in the development of novel quantum sources of light as well as for ultrafast and quantum optical signal processing circuitry~\cite{Chang2014}. Here we report very large Kerr-like nonlinearities by employing strong optical transitions of charged excitons (trions) observed in semiconducting transition metal dichalcogenides (TMDCs). By hybridising trions in monolayer MoSe$_2$ at low electron densities with a microcavity mode, we realise trion-polaritons~\cite{Dufferwiel2015, Sidler2017,Sidler2017-comment} exhibiting significant energy shifts at very small photon fluxes due to phase space filling. Most notably, the strong trion-polariton nonlinearity is found to be 10 to 1000 larger than in other polariton systems~\cite{Walker2015, Ferrier2011, Cuevas2018}, including neutral exciton-polaritons in TMDCs. Furthermore it exceeds by factors of $\sim10^3-10^5$ the magnitude of Kerr nonlinearity in bare TMDCs, graphene and other widely used optical materials (e.g Si, AlGaAs etc) in weak light-matter coupling regimes. The results are in good agreement with a theory which accounts for the composite nature of excitons and trions and deviation of their statistics from that of ideal bosons and fermions. This work opens a new highly nonlinear system for quantum optics applications enabling in principle scalability and control through nano-engineering of van der Waals heterostructures.}

Strong optical Kerr nonlinearity can be realised in photonic systems where light is resonantly coupled with optical transitions in semiconductor quantum dots~\cite{Imamoglu2014,Sun2018}, single molecules~\cite{Maser2016}  and Rydberg atoms~\cite{Volz2014}; the scalability and integration into photonic devices, nevertheless, remains a significant challenge. These obstacles can be overcome by employing hybrid 2D exciton-photon (2D polariton) systems~\cite{bookBEC, Sanvitto2016}, enabling giant effective photon-photon interactions as well as scalability and ultrafast response~\cite{Walker2015, Ferrier2011, Rodriguez2016, Cuevas2018}. So far nonlinear polaritons have been investigated in neutral exciton-polariton platform based on GaAs quantum wells and at low temperatures (4-70 K). Interaction-based effects such as polariton Bose-Einstein condensation and superfluidity~\cite{bookBEC}, solitons~\cite{Sich2016}, quantum emission~\cite{Cuevas2018, munoz2017} as well as polariton transistors/switches~\cite{Ballarini2013,Sturm2014} have been reported.

Recently, layered materials such as graphene and transition metal dichalcogenides (TMDCs)~\cite{RevModPhys.90.021001} have arisen as very promising optically active 2D media offering compatibility and ease of integration with various nanophotonic devices~\cite{Schneider2018}. Optical bistability and regenerative oscillations have been demonstrated in hybrid Si-graphene microcavities~\cite{Gu2012}. In contrast to graphene, monolayers of TMDCs exhibit Wannier-Mott excitons with a huge binding energy of about 200-500~meV and large oscillator strength~\cite{Mak2010}. This enables polariton formation in photonic structures with just a single TMDC monolayer~\cite{Dufferwiel2015} and at room temperature~\cite{Liu2014, Zhang2018}, offering a major advantage over other semiconductor polariton platforms (such as GaAs). Importantly, the strong Coulomb interactions give rise to very robust 2D trions (charged excitons) in TMDCs. The large oscillator strength of trions enables formation of well-resolved trion-polariton resonances at relatively small electron density~\cite{Dufferwiel2015,Sidler2017,Sidler2017-comment}, which, as we show here, leads to a pronounced phase space filling effect enabling nonlinearity of one to three orders of magnitude bigger than that of neutral exciton-polaritons in both GaAs and TMDC platforms. Furthermore, the nonlinear refractive index ($n_2$) per single TMDC monolayer due to trion-polaritons is observed to be three to five orders of magnitude greater than in bare 2D TMDC materials and graphene studied in the weak light-matter coupling regime. We also probe nonlinearities due to neutral exciton-polaritons, which are observed to decrease by more than an order of magnitude with power. Such a result is explained by three-exciton and possibly trion-mediated exciton-exciton scattering processes.
\begin{figure*}[t!]
\centering
\includegraphics[width=\textwidth]{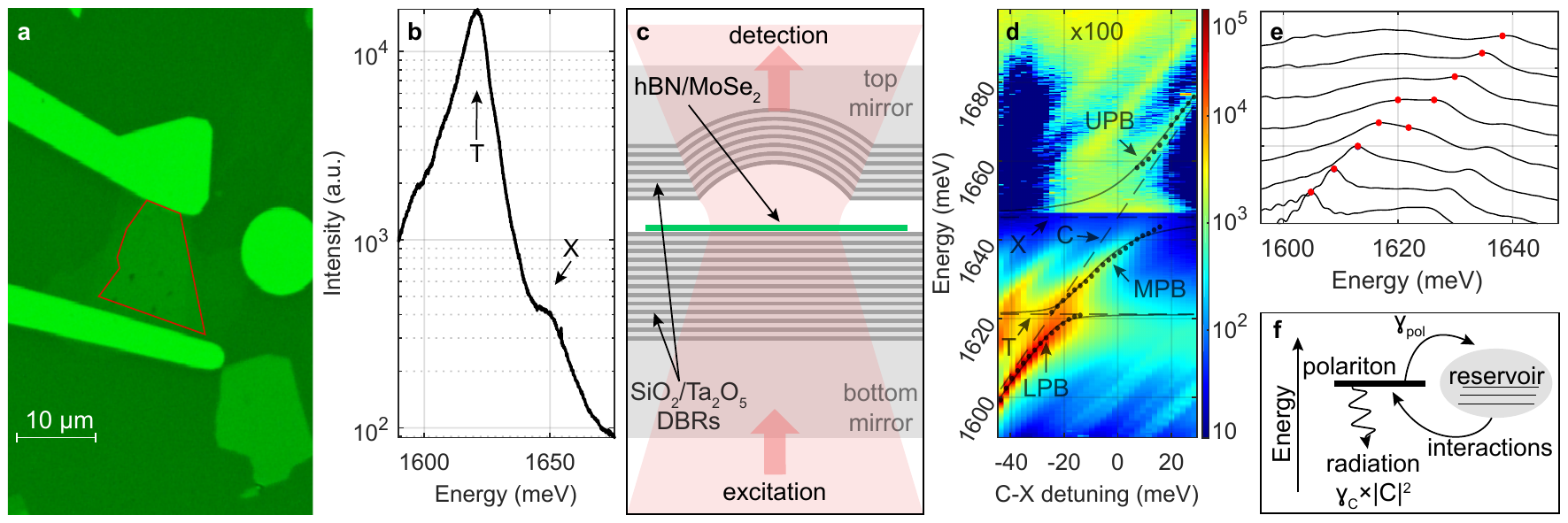}
\caption{\textbf{a} Optical microscope image of the MoSe$_2$ flake, highlighted with a red border, encapsulated in hBN and clamped with two golden contacts on top of the bottom mirror. \textbf{b} PL emission of the flake under off-resonant excitation with exciton (X) and trion (T) peaks highlighted. The observed order of magnitude difference for the intensity of the PL of the trion relative to the exciton arises from fast neutral exciton to trion states relaxation processes due to scattering with phonons and electrons. \textbf{c} Scheme of experiment for resonant excitation of the open cavity in the transmission geometry. \textbf{d} PL emission as a function of the detuning between bare cavity $LG_{00}$ mode (C) and exciton (X) plotted in logarithmic colour scale; black points (dots) are experimental peak positions of the polariton resonances extracted from the fitting of experimental spectra at each detuning with a Gaussian; solid lines are theoretical polariton branches. Higher order transverse LG modes are also observed on the left and right-hand side of $LG_{00}$ polariton states. \textbf{e} Normalised PL spectra around trion resonance, showing anticrossing behaviour.  \textbf{f} Schematic showing emission and absorption of driven polariton resonance and interactions with the reservoir of trions and excitons.}
\label{fig:intro}
\end{figure*}

In our experiment a monolayer of MoSe$_2$  (Fig.~\ref{fig:intro}a) was placed into an open-access microcavity (MC) (see Fig.~\ref{fig:intro}c~\cite{Giriunas2018}) where Laguerre-Gauss (LG) photonic modes strongly couple to exciton and trion states forming polaritons. Fig.~\ref{fig:intro}b shows the spectrum of PL emission from the monolayer. The strong peak at lower energy $\sim 1.62$~eV arises from the trion (T) emission due to the natural doping in the sample, while the weaker peak at $\sim 1.65$~eV corresponds to neutral (X) excitons. To characterise the polaritons the MC was excited with a laser at 1.95~eV and polariton emission was recorded as a function of the detuning between the excitons and $\LGOO$ mode. In Fig.~\ref{fig:intro}d the lower, middle and upper polariton branches (labelled LPB, MPB and UPB) are observed due to the strong coupling of the photon mode with T and X, respectively. The peak polariton positions (black dots in Figs.~\ref{fig:intro}d, e) were fitted using a model of three coupled oscillators (see SM, secs.A,C, for the details).

In order to probe the polariton nonlinearity the polaritons were excited resonantly using a laser with 100-fs pulse duration. The excited polaritons then decay via photon emission through the DBRs with a rate $\gamma_C\cdot\left|C\right|^2$ given by the cavity mode linewidth $\gamma_C\sim0.4$~meV ($\left|C\right|^2$ is the photonic fraction) or are absorbed in the 2D material with a rate given by the polariton linewidth $\gamma_{pol}$ ($\gamma_{pol}\sim$~3-5~meV~$\gg \gamma_C \cdot \left|C\right|^2$) due to scattering with exciton disorder resulting in creation of an exciton-trion reservoir (Fig.~\ref{fig:intro}f)~\cite{Walker2017, Giriunas2018}. Measurements of the intensity of transmitted light enable us to estimate accurately the polariton density excited inside the MC and the density of the reservoir. Monitoring the frequency shift of the polariton resonance with increasing pump power, we observe the influence of interaction between polaritons and the exciton-trion reservoir (Fig.~\ref{fig:intro}f), and can extract the strength of exciton- and trion-based nonlinearity.
\begin{figure}
\centering
\includegraphics[width=\columnwidth]{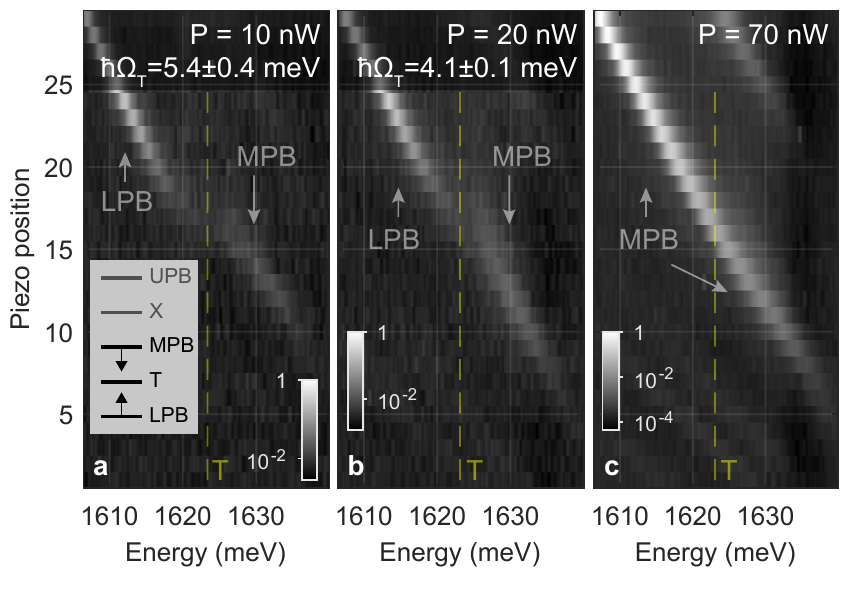}
\caption{\textbf{a-c} Cavity scan of $\LGOO$ mode across the trion resonance for different pump powers: 10, 20, and 70 nW, respectively. The peak energy of the pump laser is fixed at $\sim 1.62$~eV. Pseudo-colour scale is logarithmic. Estimates for the trion Rabi splittings are obtained by fitting extracted peak positions for each piezo step with a coupled oscillator model. {\textbf{Inset}} Schematic diagram showing how the energies MPB and LPB renormalise in the case of phase space filling effect leading to reduction of Rabi-splitting between trion level and the bare photon mode. }
\label{fig:CTscan}
\end{figure}

We first present the results for interacting trion-polaritons. In this experiment the laser energy is fixed at approximately the trion level, $E_p=1.62$~eV and the transmission spectrum is recorded as the energy of the photon $\LGOO$ mode is scanned through the trion resonance. Since the linewidth of the pulsed laser ($\sim 15$~meV) is significantly larger than the observed photon-trion Rabi-splitting of $\hbar\Omega_{T} \approx 5.8$~meV, it was possible to inject a similar number of polaritons in the vicinity of the trion resonance for each cavity mode position. Excitation of the UPB is negligible since it is located at energies more than $\sim 30$ meV above the trion-polariton states.

Fig.~\ref{fig:CTscan} shows the results of three cavity scans performed for different pump powers. At the lowest pulse power, 10~nW (Fig.~\ref{fig:CTscan}a), an anti-crossing between the cavity mode and the trion level is clearly observed, as in the case under the non-resonant excitation. With an increase of the pump power the photon-trion Rabi-splitting is reduced (Fig.~\ref{fig:CTscan}b) leading to the the blueshift and redshift of the LPB and MPB states in the vicinity of trion resonance (Fig.~\ref{fig:CTscan}a, inset), respectively. At $P=70$~nW no anticrossing is observed (Fig.~\ref{fig:CTscan}c), with MPB and LPB merging together and forming a single polariton branch.
\begin{figure}[t]
\centering
\includegraphics[width=\columnwidth]{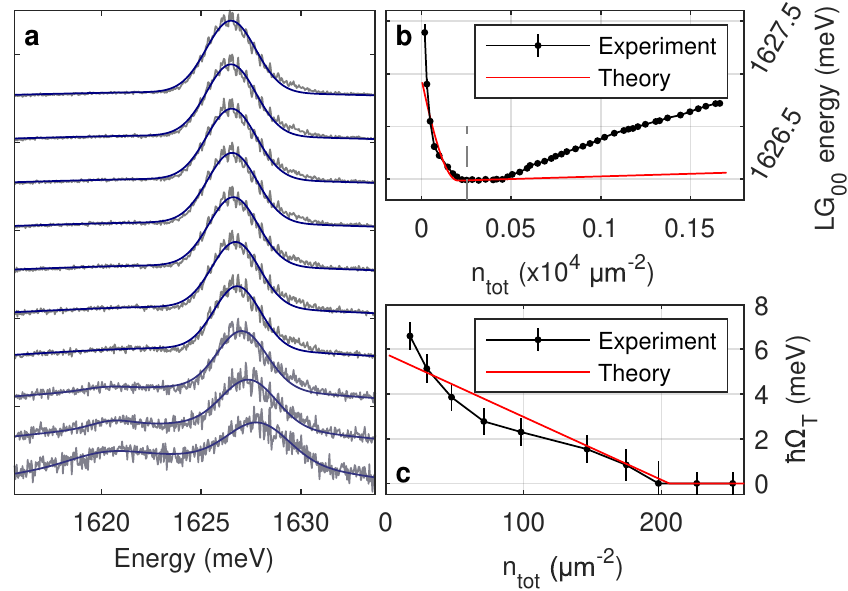}
\caption{$\delta_{\mathrm{C-X}}=-15.4$~meV. \textbf{a} Spectra of the MPB and LPB  for the excitation power from 5 to 50 nW (power increases from bottom to top). The MPB and LPB peaks are fitted with Gaussians. The fitting of the exciton-like LPB peak (lower energy peak) is performed for the first 4 powers, since it becomes broad and weak at higher densities and the fitting procedure is not reliable. \textbf{b} Extracted peak positions of the MPB mode vs estimated total exciton-trion density. \textbf{c} Trion-polariton Rabi-splitting vs exciton-trion density. Red solid curves in \textbf{b} and \textbf{c} correspond to the theoretical modelling results. The error bars ($95\%$ CI) are estimated taking into account the random error in determination of the peak positions in \textbf{a} as well as possible systematic error ($\Delta\Omega_T\sim0.5$ meV) due to the uncertainty of the fitting parameters in the coupled oscillators model.}
\label{fig:Tdetail}
\end{figure}
To quantify the strength of the nonlinearity, we recorded the spectra of the transmitted light as a function of pump power at a fixed photon mode energy near the trion resonance (Fig.~\ref{fig:Tdetail}a), at a fixed photon-exciton detuning $\delta_{\mathrm{C-X}}=-15.4$~meV. In Fig.~\ref{fig:Tdetail}a two peaks are observed at low power of 5 nW; the MPB peak has a higher intensity due to the higher photonic fraction. With increase of the pulse power the MPB peak exhibits a fast redshift. The intensity of the trion-like LPB quickly reduces down to zero with power due to reduction of its photon fraction as the strong trion-photon coupling collapses.

The collapse of the strong coupling is driven by the population of both polaritons inside the cavity and reservoir determining the total density of excited excitons and trions $n_{tot}$ (see SM, sec.~A2, for estimation of $n_{tot}$). The reservoir may consist of localised or dark exciton/trion states as well as high-momenta trions degenerate with trion-polariton resonances~\cite{Walker2017, Giriunas2018}. For the case of resonantly driven trion-polaritons $n_{tot}$ is dominated mostly by the trion density $n_T$  ($n_{tot} \approx n_{T}$), since the peak of the exciton density of states is blue detuned by $\sim30$~meV from the trion.

Fig.~\ref{fig:Tdetail}b shows that the MPB energy shifts rapidly to the red wavelengths with $n_{tot}$ at densities $n_{tot}<2\cdot10^2~\iums$ and then exhibits a plateau corresponding to the quenching of trion-photon coupling, followed by a gradual increase. From the redshift of the MPB branch in Fig.~\ref{fig:Tdetail}b we plot (see SM, sec.~A4) the value of the Rabi-splitting $\hbar \Omega_T$ vs $n_{tot}$ in Fig.~\ref{fig:Tdetail}c. The collapse of trion-photon coupling occurs at a small density of $n_{tot}\sim2\cdot10^2~\iums$ very close to half the estimated density of free electrons $n_{e}/2\approx 2\cdot10^2~\iums$ (see SM, sec.~C2). The effective strength of the trion-polariton nonlinearity $\beta_T^{\textrm{eff}}$ responsible for the quenching of strong coupling (and large energy shifts) is defined as $\beta_T^{\textrm{eff}} = -d(\hbar \Omega_T) / d n_{tot}$~\cite{Brichkin2011}. From Fig.~\ref{fig:Tdetail}c we deduce an average value of $\beta_T^{\textrm{eff}} \simeq 37 \pm 3~\mu\textrm{eV}\cdot\mu\textrm{m}^2$, which is $\sim 10$ times larger than that for neutral exciton-polaritons in GaAs(see SM, sec.~A4)~\cite{Brichkin2011, Walker2015}.
We also deduce the effective nonlinear refractive index $n_{2\mathrm{(MoSe_2)}}$ ($n_{2\mathrm{(MoSe_2)}}\sim \beta_T^{\textrm{eff}}$) per single flake associated with strong trion-photon coupling to be around $\sim10^{-10}$ m$^2$/W (see SM, sec.~A7). This value is 3-5 orders of magnitude greater than $n_2$ coefficient of 2D TMDC materials and graphene studied in the weak light-matter coupling regime (see SM, sec.~A7).  

To explain the observed result we account for phase space filling effects which become important at increasing excited trion density (see SM, sec.~C). This leads to the quenching of the collective trion oscillator strength: as more and more trions are excited in the system, the extra injected photons have less electrons and trions to couple to and to form trion-polaritons. As a result the collapse of strong photon-trion coupling occurs at the density of excited trions equal to half the density of available free electrons, $n_T \approx n_e/2$. The results of the theoretical modelling (see Methods) are shown by the red solid curves in Figs.~\ref{fig:Tdetail}b and c reproducing the experimental \textit{red} shift of $E_{\mathrm{MPB}}$ and the corresponding reduction of $\hbar \Omega_T$. Overall, the trion Rabi-splitting can be approximated as $\Omega_T (n_T) = \Omega_T (0) ( 1 - 2 n_T/n_e)$, and the value of theoretical nonlinearity is given by $\beta_T^{\textrm{eff}} = 2\hbar\Omega_T(0) / n_e \approx 30 ~\mu\textrm{eV}\cdot\mu\textrm{m}^2$ for $n_e \approx 400$ $\mu$m$^{-2}$ and $\hbar \Omega_T(0) \approx 5.8$~meV  (see SM, sec.~C5) in agreement with the experimental value of $37 \pm 3~\mu\textrm{eV}\cdot\mu\textrm{m}^2$. It is the low electron density and high oscillator strength per single trion (large $\hbar \Omega_T(0)$), that lead to the high value of trion-polariton nonlinearity.



\begin{figure}
\centering
\includegraphics[width=\columnwidth]{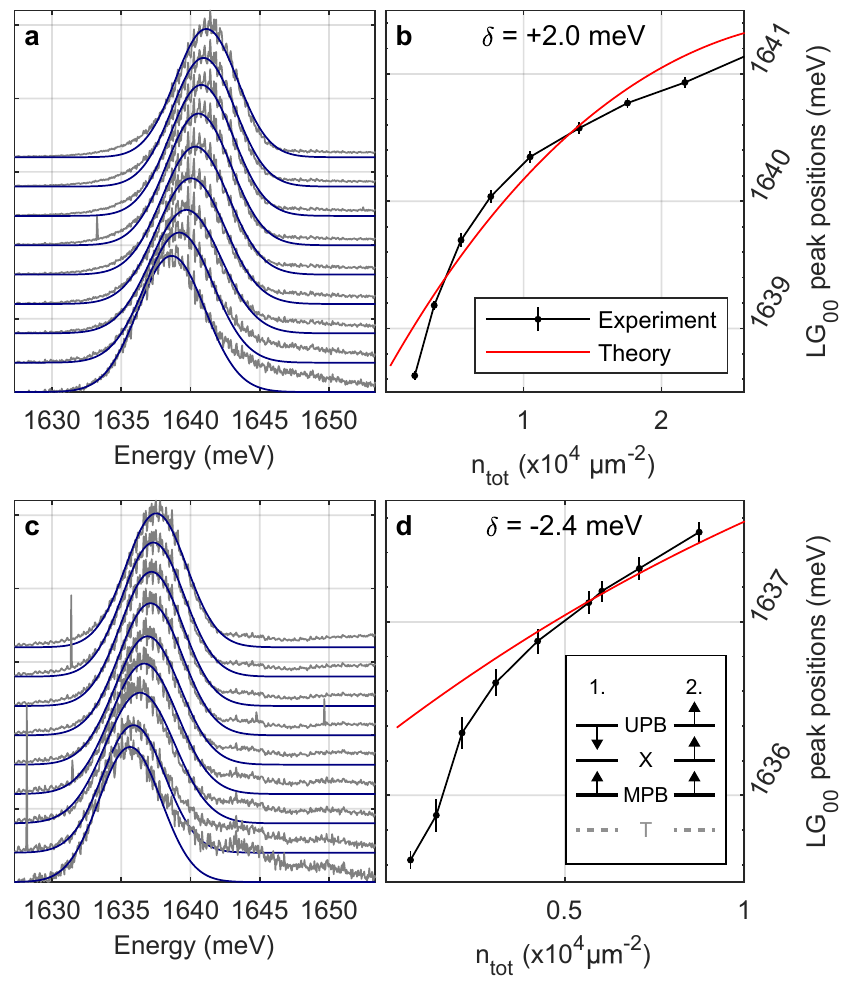}
\caption{\textbf{a,b} $\delta_{C-X}=+2.0$~meV. \textbf{a} Spectra of the MPB $\LGOO$ mode for the pump powers from 1 to 9 $\mu$W (power increases from bottom to top). \textbf{b} Extracted peak positions of the mode vs. exciton density. \textbf{c,d} $\delta_{C-X}=-2.4$~meV. \textbf{c} Spectra of the MPB $\LGOO$ mode for the pump powers from 1 to 9 $\mu$W (power increases from bottom to top). \textbf{d} Extracted peak positions of the mode vs. estimated exciton density. The error bars ($95\%$ CI ) in \textbf{b-d} are deduced from the fitting procedure in \textbf{a-c}. Red solid curves correspond to the theoretical modelling results. {\textbf{Inset}} Schematic diagrams showing how the MPB and UPB shift in the case of phase space filling (Mechanism 1 leading to reduction of Rabi-splitting between neutral exciton level and the bare photon mode) and the neutral exciton blueshift (Mechanism 2).}
\label{fig:Xdetail}
\end{figure}

Next we study the neutral exciton-polariton nonlinearity, which may arise from: (1) the reduction of exciton-photon Rabi-splitting $\hbar \Omega_X$, and/or (2) blueshift of the neutral exciton level $E_X$ \cite{Brichkin2011}. Both mechanisms should lead to blueshift of the MPB. The blueshift of the MPB peak associated with the neutral exciton-polariton nonlinearity is observed in Fig.~\ref{fig:Tdetail}b for $\delta_{\mathrm{C-X}}=-15.4$~meV at $n_{tot}>5\cdot10^2~\iums$ above the threshold of strong trion-photon coupling collapse. We further studied neutral exciton-polariton nonlinearity for several photon-exciton detunings in the range from $+8.8$~meV to $-2.4$~meV, where the trion fraction is negligible ($\sim 3 \%$ or less; see Table~II in SM). The central laser frequency was shifted to be in resonance with the MPB in each case. Substantial blueshifts of MPB mode of the order of 2~meV are observed at much higher excitation power in the range from 1 to 9 $\mu$W (see Fig.~\ref{fig:Xdetail}a and c). The MPB energy depends sublinearly on the total exciton/trion density $n_{tot}$ as shown in Fig.~\ref{fig:Xdetail}b and d. In this power range the density of excited neutral excitons $n_X$ is much higher than that of trions and $n_{tot}$ is dominated mostly by neutral excitons ($n_{tot} \gg n_e$, $n_{tot}\approx n_{X}$). 

\begin{figure}
\centering
\includegraphics[width=\columnwidth]{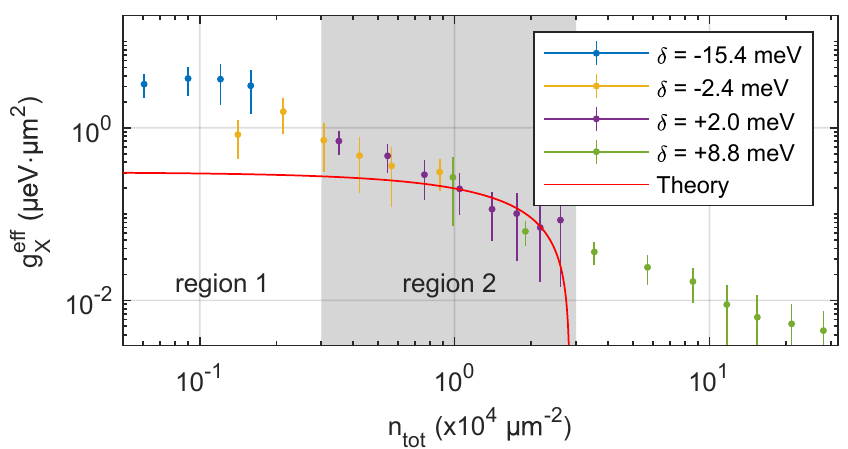}
\caption{The experimental effective interaction constant $g_X^{\textrm{eff}}$ as a function of the estimated exciton density, $n_{tot}$. The $g_X^{\textrm{eff}}$ is deduced from the energy blueshift of MPB (see SM, sec.~A5). The data correspond to four different cavity-exciton detunings ($\delta$): $+8.8$~meV (olive), $+2.0$~meV (purple), $-2.4$~meV (orange), and $-15.4$~meV (blue). The error bars ($95\%$ CI) are deduced taking into account errors in determining the MPB peak positions at each pump power (exciton density). The red solid curve corresponds to the theoretical values.}
\label{fig:Xbeta}
\end{figure}

As we discuss in sec.~A5 of SM at intermediate densities the optical nonlinearity arises mainly from the exciton blueshift (Mechanism 2)~\cite{Brichkin2011}, which is characterised by the parameter $g_X^{\textrm{eff}} = d E_X / d n_{tot}$. The $g_X^{\textrm{eff}}$ is expected to be constant in a system where only pair exciton-exciton interactions are important~\cite{Brichkin2011}. By contrast,  Fig.~\ref{fig:Xbeta} shows that the experimental $g_X^{\textrm{eff}}$ decreases with $n_{tot}$ from $\simeq$ 2-5 $\mu\textrm{eV}\cdot\mu\textrm{m}^2$ at $n_{tot}\sim 10^{3}$ $\mu$m$^{-2}$ to  $\simeq 0.01 ~\mu\textrm{eV}\cdot\mu\textrm{m}^2$ at $n_{tot}\sim 10^{5}$ $\mu$m$^{-2}$, which suggests the importance of higher order exciton-exciton interactions. Overall, in TMDC the neutral exciton-polariton nonlinearity is one to three orders of magnitude lower than that of trion-polaritons. 
The lower values $\simeq 0.05~\mu\textrm{eV}\cdot\mu\textrm{m}^2$ are similar to the values reported in WS$_2$ waveguide structures, where only very high excitation powers were used~\cite{Barachati2018}.

To describe the observed neutral exciton-nonlinearity we developed a model taking into account the two- and three-exciton exchange processes~\cite{Tassone1999,Shahnazaryan2017} (see Methods and SM, sec.~B). At $3\cdot 10^3 < n_{tot} < 3\cdot10^4$ $\mu$m$^{-2}$ within an experimental error there is agreement between theory and experiment (Fig.~\ref{fig:Xbeta}, region 2). \textcolor{black}{The values of $g_X^{\textrm{eff}}$ in exciton density region 2 are also in agreement with the values of the exciton-exciton interaction parameter in monolayer WSe$_2$ characterising excitation induced exciton broadening in the same density range \cite{Moody2015}.}
The theory also qualitatively reproduces the sublinear shifts in Fig.~\ref{fig:Xdetail}b and d. At $n_{tot} > 3\cdot 10^4$  $\mu$m$^{-2}$ our theory is not applicable anymore, since in this case higher order nonlinearities should be taken into account. By contrast, at $n_{tot} < 3\cdot10^3$ $\mu$m$^{-2}$ the model accounting only for exciton-exciton interactions results in the very weak theoretical blueshift of $E_{\mathrm{MPB}}$, much smaller than that observed in the experiment in Fig.~\ref{fig:Tdetail}b. This is also reflected in Fig.~\ref{fig:Xbeta}, where at $n_{tot} < 3\cdot10^3$ $\mu$m$^{-2}$ (region 1) the theoretical $g_X^{\textrm{th}}\sim 0.3 ~\mu\textrm{eV}\cdot\mu\textrm{m}^2$ is observed to be below the corresponding experimental values $g_X^{\textrm{eff}} \sim 0.5-3 ~\mu\textrm{eV}\cdot\mu\textrm{m}^2$. 
Such a discrepancy indicates that at these exciton densities the trion-mediated exciton-exciton interactions, which are characterised by increased scattering cross-section and number of exchange processes, might play an important role in the observed large polariton blueshift (see SM, sec.~C6).

In conclusion we observed giant Kerr nonlinearity associated with trion-polaritons in TMDC system due to phase space filling of trion states. The strong variation of neutral-exciton polariton nonlinearity with density is attributed to higher order exciton-exciton and trion-exciton interactions.  Notably, the trion-polariton nonlinearity is of the same order as that observed in microcavities with a single quantum dot~\cite{Imamoglu2014,Sun2018}, where strong renormalisation of Rabi-splitting ($\sim 100~\mu$eV) occurs at a single photon level~\cite{Imamoglu2014}. \textcolor{black}{Therefore, as we elaborate theoretically in SM, sec.D} polariton quantum effects~\cite{Chang2014} can potentially also be realised in high Q microcavities with embedded high quality homogeneous TMDC samples~\cite{Cadiz2017}, where very narrow polariton resonances with a linewidth given just by the cavity mode lifetime could be achieved. Our work paves the way towards development of \textit{scalable} active nanophotonic devices based on 2D materials (where the exciton level is the same across large areas in contrast to 0D quantum dots) utilising the polariton nonlinearity for control of light by light, potentially at quantum level.

\section{Methods}
\subsection{Experimental technique}

In our experiment a MoSe$_2$ monolayer is covered with a monolayer of hexagonal boron nitride (hBN) to protect MoSe$_2$ from contamination. The heterostructure of hBN/MoSe$_2$ was fabricated using standard mechanical exfoliation technique and dry transfer methods. To form a microcavity the hBN/MoSe$_2$ structure was positioned on top of a flat distributed Bragg reflector (DBR) (bottom mirror) consisting of 13 pairs SiO$_2$/Ta$_2$O$_5$ quarter-wave layers. The top mirror is a 13-pair DBR deposited on a hemispherical surface, which is fabricated using focused ion beam milling. The two mirrors were then aligned and brought into a close proximity to each other (the distance between the mirrors is set to $\sim 1~\um$) using piezo nano-positioners forming an ``open-cavity'' system (see Fig.~\ref{fig:intro}c~\cite{Giriunas2018}) with the resulting formation of discrete microcavity Laguerre-Gauss (LG) photonic modes. The open cavity system was positioned into bath cryostat at temperature of 4~K. For resonant excitation we employed 100~fs laser pulses with repetition rate 1~kHz. These were obtained from the frequency-doubled output of an optical parametric amplifier (Light-Conversion TOPAS) pumped by the 800~nm pulses from a Ti:Sapphire regenerative amplifier system (Spectra Physics Spitfire seeded by Spectra-Physics Mai-Tai Ti:Sapphire Oscillator). \textcolor{black}{The excitation beam was focused into the spot size of about $5-10$ $\mu$m on the bottom flat mirror, so that only a fraction of the incident photons couples to the highly confined $\LGOO$ cavity mode.}

\subsection{Theory methods}

To describe the observed polariton nonlinearity we begin with the conceptually more simple case of a neutral exciton-polariton, where the trion fraction is negligible (Fig.~\ref{fig:Xdetail}). Aiming to describe the influence of the exciton density ($n_X$) increase, we note that the Rabi frequency $\Omega_{X}$ corresponds to the low intensity value $\Omega_{X}^{(0)}$ only for the case when excitons are ideal bosons described by a creation operator $\hat{X}^\dagger_{\mathbf{k}}$. The deviation of statistics coming from the Pauli principle leads to their changed commutation relations, $[\hat{X}_{\mathbf{k}}, \hat{X}^\dagger_{\mathbf{k}'}] = \delta_{\mathbf{k},\mathbf{k}'} - D_{\mathbf{k},\mathbf{k}'}$~[M1], where $D_{\mathbf{k},\mathbf{k}'}$ is a non-bosonicity parameter. In the analysis we consider density-dependent Rabi-splitting $\hbar \Omega_{X}(n_X)$ and nonlinear shift of the exciton level $E_X(n_X)$ (see Supp. Mat., sec.~B, for the details). The modified coupling reads
\begin{align}
\label{eq:Rabi_X}
\Omega_{X}(n_X) = \Omega_{X}(0) \left(1 - \frac{8\pi}{7} n_X a_B^2 + \frac{384 \pi^2}{455} n_X^2 a_B^4 \right),
\end{align}
where $a_B$ is an exciton Bohr radius in TMDC. The nonlinear shift of the exciton energy arising from the Coulomb scattering and deviation of exciton statistics from that of ideal bosons, reads
\begin{align}
\label{eq:E_X}
&E_{X}(n_X) = E_{X}(0) + \frac{8}{\pi} \frac{e^2}{4\pi \epsilon_0 \kappa a_B} \mathcal{I}_4(r_0) n_X a_B^2 \\ \notag &- \frac{128}{5} \frac{e^2}{4\pi\epsilon_0 \kappa a_B} \Big[5 \mathcal{I}_6(r_0) -2 \mathcal{I}_4(r_0) \Big] n_X^2 a_B^4,
\end{align}
where $e$ is an electron charge, $\epsilon_0$ is vacuum permittivity, $\kappa$ denotes average dielectric permittivity, $r_0$ is a screening parameter, and $\mathcal{I}_{4,6}(r_0)$ are dimensionless integrals. The linear ($\propto n_X$) and quadratic terms ($\propto n_X^2$) in Eq.~\eqref{eq:Rabi_X} and Eq.~\eqref{eq:E_X} take into account the two- and three-exciton exchange processes~\cite{Tassone1999,Shahnazaryan2017}, respectively. 

We apply the above theory to explain the nonlinear MPB blueshift when trion contribution can be omitted. To do so, the exciton-photon Hamiltonian is diagonalised accounting both for $E_{X}(n_X)$ and $\Omega_{X}(n_X)$, and energy for the lower polariton branch is considered~\cite{Brichkin2011}. The theoretical results are shown in Fig. \ref{fig:Xdetail}(b,d) by the red solid curves. Performing the variational procedure to calculate the properties of excitons and matching the binding energy to experimental value, we set the exciton Bohr radius to $a_B = 0.85$~nm~[M2] and qualitatively describe the sublinear dependence of polariton blueshift (see SM., sec.~B). By differentiating Eqs. \eqref{eq:Rabi_X} and \eqref{eq:E_X} over $n_X$ we can obtain the corresponding theoretical $\beta_X^{\textrm{th}}$ and $g_X^{\textrm{th}}$ parameters, which characterize the neutral exciton-polariton nonlinearity due to reduction of exciton-polariton Rabi-splitting and the exciton level blueshift, respectively (see SM, sec.A5).

Gaining the knowledge from the exciton case, we consider the trion-dominated regime (Figs. \ref{fig:CTscan} and \ref{fig:Tdetail}). The excitation process then corresponds to the creation of a trion from a free electron~[M3, M4]. We account for phase space filling effects which become important at increasing excited trion density (see SM, sec.~C). This leads to the quenching of the collective trion oscillator strength. Taking into account the deviation of statistics for trions from that of ideal fermions~[M5] we calculate the influence of phase space filling on the trion Rabi frequency $\Omega_{T}(n_T,\lambda_{1,2})$ as a function of density and variational parameters for the trion wavefunction $\lambda_{1,2}$. To describe Figs.~\ref{fig:Tdetail}b,c we perform the diagonalisation of the full photon-trion-exciton system with nonlinear contributions to both trion and exciton modes. We use $\lambda_1 = 0.86$~nm, $\lambda_2 = 2.54$~nm, obtained by the variational procedure, in line with computational trion studies in MoSe$_2$~[M6]. The results are shown by the red solid curves in Figs.~\ref{fig:Tdetail}b,c reproducing the experimental red shift of $E_{\mathrm{MPB}}$ and reduction of $\Omega_{T}$ .

\section{Data availability}
The data that support the findings of this study are available from the corresponding author on reasonable request.

\section{Code availability}
All the detailed theoretical derivations are presented in the Supplementary materials and can be easily implemented in, for example, Mathematica package.

\section{Acknowledgements}
MS, RPAE, PMW, MSS and DNK acknowledge the support from the EPSRC grant EP/N031776/1; MSS and AIT acknowledge the support from the EPSRC grant EP/M012727/1; MS, AIT, DNK acknowledge the support from the EPSRC grant EP/P026850/1; VS, MSS, IAS and DNK acknowledge support from the mega-grant No. 14.Y26.31.0015 of the Ministry of Education and Science of the Russian Federation. VS and IAS acknowledge support from goszadanie no 3.2614.2017/4.6 of the Ministry of Education and Science of the Russian Federation and Icelandic research fund, grant No. 163082-051.

\section{Author contributions}
RPAE and MS performed the experiments and analysed the data; FW and AC provided the samples; PMW designed the cavity structure; OK, VS, and IAS provided the theoretical explanation; DNK oversaw the experiment; MS, OK, PMW, MSS, AIT, and DNK wrote the paper.

\section{Competing interests}
The authors declare no competing interests.

\section{Materials \& Correspondence}
Correspondence and material requests should be addressed to Dimitrii Krizhanovskii at d.krizhanovskii@sheffield.ac.uk. Correspondence regarding the theoretical description should be addressed to Oleksandr Kyriienko at kyriienko@ukr.net.

\centering
\textbf{REFERENCES [METHODS]}
\begin{UR}  
\item Combescot, M., Betbeder-Matibet, O. \& Dubin, F. The many-body physics of composite bosons. Physics Reports 463, 215–320 (2008).
\item Kyl\"{a}np\"{a}\"{a}, I. \& Komsa, H.-P. Binding energies of exciton complexes in transition metal dichalcogenide monolayers and  effect  of  dielectric  environment. Phys.  Rev.  B 92, 205418 (2015).
\item Rapaport, R., Cohen, E., Ron, A., Linder, E. \& Pfeiffer, L. N. Negatively charged polaritons in a semiconductor microcavity. Phys. Rev. B 63, 235310 (2001).
\item Shiau,  S.-Y.,  Combescot,  M.  \&  Chang,  Y.-C.  Trion ground  state,  excited  states,  and  absorption  spectrum using  electron-exciton  basis. Phys.  Rev.  B 86, 115210 (2012).
\item Combescot, M. \& Betbeder-Matibet, O. General Many-Body Formalism for Composite Quantum Particles. Phys. Rev. Lett. 104, 206404 (2010).
\item Kidd, D. W., Zhang, D. K. \& Varga, K. Binding energies and  structures  of  two-dimensional  excitonic  complexes in  transition  metal  dichalcogenides. Phys.  Rev.  B 93, 125423 (2016).
\end{UR}


\begin{thebibliography}{99}
\bibitem{Sidler2017-comment} In Ref. [3] trion-polaritons have been interpreted in terms of Fermi-polaron polaritons. However, as we discuss in SM (sec. C) in our sample the Fermi wave vector of electrons is less than the inverse Bohr radius of excitons, and thus the trion formalism is the appropriate description.
\end{thebibliography}
\end{document}


\title{Supplemental Material: Highly nonlinear trion-polaritons in a monolayer semiconductor}

\author{R. P. A. Emmanuele}
\affiliation{Department of Physics and Astronomy, The University of Sheffield, Sheffield, S3~7RH, United Kingdom}

\author{M. Sich}
\affiliation{Department of Physics and Astronomy, The University of Sheffield, Sheffield, S3~7RH, United Kingdom}

\author{O. Kyriienko}
\affiliation{Department of Physics and Astronomy, University of Exeter, Stocker Road, Exeter EX4 4QL, UK}
\affiliation{Department of Nanophotonics and Metamaterials, ITMO University, St. Petersburg, 197101, Russia}

\author{V. Shahnazaryan}
\affiliation{Department of Nanophotonics and Metamaterials, ITMO University, St. Petersburg, 197101, Russia}

\author{F. Withers}
\affiliation{College of Engineering, Mathematics and Physical Sciences, University of Exeter, Exeter, EX4 4QF, United Kingdom}

\author{A. Catanzaro}
\affiliation{Department of Physics and Astronomy, The University of Sheffield, Sheffield, S3~7RH, United Kingdom}

\author{P.~M. Walker}
\affiliation{Department of Physics and Astronomy, The University of Sheffield, Sheffield, S3~7RH, United Kingdom}

\author{M.~S. Skolnick}
\affiliation{Department of Physics and Astronomy, The University of Sheffield, Sheffield, S3~7RH, United Kingdom}
\affiliation{Department of Nanophotonics and Metamaterials, ITMO University, St. Petersburg, 197101, Russia}

\author{A.~I. Tartakovskii}
\affiliation{Department of Physics and Astronomy, The University of Sheffield, Sheffield, S3~7RH, United Kingdom}

\author{I.~A. Shelykh}
\affiliation{Science Institute, University of Iceland, Dunhagi-3, IS-107 Reykjavik, Iceland}
\affiliation{Department of Nanophotonics and Metamaterials, ITMO University, St. Petersburg, 197101, Russia}

\author{D.~N. Krizhanovskii}
\affiliation{Department of Physics and Astronomy, The University of Sheffield, Sheffield, S3~7RH, United Kingdom}
\affiliation{Department of Nanophotonics and Metamaterials, ITMO University, St. Petersburg, 197101, Russia}
\email{d.krizhanovskii@sheffield.ac.uk}

\date{\today}

\begin{abstract}
Here we describe the experimental details (sec. A), and the microscopic theory for exciton-polaritons (sec. B) and trion-polaritons (sec. C) which were used for the theoretical modelling. In Sec. D we discuss possible quantum effect with trion-polaritons in transition metal dichalcogenides materials.
\end{abstract}

\maketitle

\tableofcontents


\section{Experimental details}

\subsection{Parameters characterising strong coupling between photons and exciton and trion }

\begin{figure}
\centering
\includegraphics[width=0.6\columnwidth]{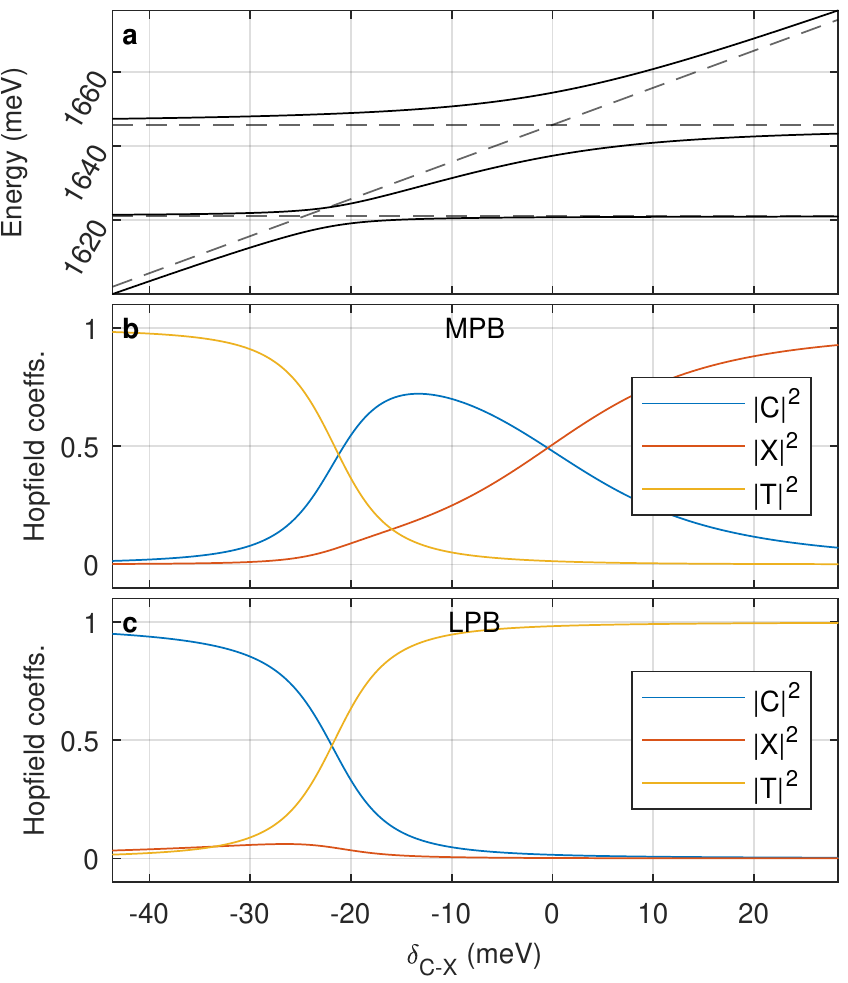}
\caption{\textbf{a} Theoretical polariton energies vs the detuning between bare cavity mode and exciton (solid lines)  obtained from the fit of the experimental polariton energies in Fig.~1d of the main text. Bare cavity mode, exciton, and trion levels are shown by dashed lines. \textbf{b,c} $\left|C\right|^2, \left|X\right|^2, \left|T\right|^2$ Hopfield coefficients of the middle (MPB) and lower (LPB) polariton branches, respectively.}
\label{fig:Hopf3}
\end{figure}

To describe the polariton system where the photonic mode is strongly coupled to both excitons and trions we employ a model of three coupled oscillators, which can be described in a compact way by the matrix equation
\begin{equation}
    \begin{bmatrix}
    E_C(L) & \frac{1}{2}\hbar\Omega_{X} & \frac{1}{2}\hbar\Omega_{T} \\
    \frac{1}{2}\hbar\Omega_{X} & E_X & 0 \\
    \frac{1}{2}\hbar\Omega_{T} & 0 & E_T 
    \end{bmatrix} \psi = E \psi,\label{eq:model3}
\end{equation}
where $E_C(L)$, $E_X$, $E_T$ are the cavity photon, exciton, and trion energies; $\hbar\Omega_{X}$ and $\hbar\Omega_{T}$ are the cavity mode-exciton and the cavity mode-trion coupling strength. $\psi$ is a three-component basis vector. In our system we can tune $E_C$ by changing the cavity length $L$. Eigenvalues of the matrix in the left-hand side of the Eq.~(\ref{eq:model3}) give the polariton mode resonances of upper, middle, and lower polariton branches (UPB, MPB, and LPB, respectively). Their eigenvectors provide the Hopfield coefficients. We start by fitting this model to the experimental data in Fig.~1b (main text), and obtained the following values for the ground $\LGOO$ mode: $E_X =1646.0\pm0.5$~meV, $E_T=1621.2\pm0.5$~meV, $\hbar\Omega_{X}=17.2\pm0.5$~meV, and  $\hbar\Omega_{T}=5.8\pm0.5$~meV. Using these parameters we can plot polariton resonances and Hopfield coefficients for different values of cavity resonance energies $E_C$ (cavity lengths, $L$) expressed as a cavity-exciton detuning, $\delta_{C-X}$, in Fig~\ref{fig:Hopf3}. We summarise Hopfield coefficients obtained for the four experimental cavity-exciton detunings used here in Table~\ref{tbl:hopf}. Note, that these are the values for the low-density case. In the high density regime, as we discuss in the main text and the theory section, the nonlinearity quenches photon-trion coupling, and the polariton system can be described by the two-oscillator model.


\begin{table}[]
\begin{tabular}{l|l|l|l|l|l|l|l|l|l|l|l|l|}
\hline
\multicolumn{1}{|l|}{$\delta_{C-X}=$} & \multicolumn{3}{l|}{$-15.4$~meV} & \multicolumn{3}{l|}{$-2.4$~meV} & \multicolumn{3}{l|}{$+2.0$~meV} & \multicolumn{3}{l|}{$+8.8$~meV} \\ \hline
 & UPB & MPB & LPB & UPB & MPB & LPB & UPB & MPB & LPB & UPB & MPB & LPB \\ \hline
\multicolumn{1}{|l|}{$\left|C\right|^2$} & 0.169 & 0.710 & 0.121 & 0.435 & 0.545 & 0.019 & 0.561 & 0.426 & 0.013 & 0.730 & 0.262 & 0.008 \\ \hline
\multicolumn{1}{|l|}{$\left|X\right|^2$} & 0.830 & 0.157 & 0.014 & 0.561 & 0.436 & 0.002 & 0.435 & 0.563 & 0.002 & 0.266 & 0.733 & 0.001 \\ \hline
\multicolumn{1}{|l|}{$\left|T\right|^2$} & 0.002 & 0.133 & 0.866 & 0.003 & 0.018 & 0.979 & 0.004 & 0.011 & 0.985 & 0.004 & 0.005 & 0.991 \\ \hline
\end{tabular}
\caption{Fitted values of the Hopfield coefficients for UPB, MPB, and LPB branches for the four experimental values of cavity-exciton detunings.}\label{tbl:hopf}
\end{table}

\begin{figure}
    \centering
    \includegraphics[width=0.6\columnwidth]{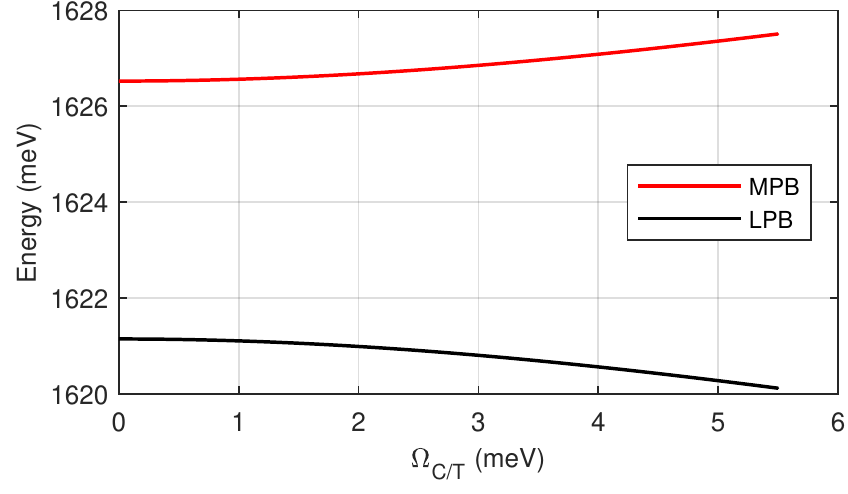}
    \caption{$\LGOO$ peak positions as function of $\hbar\Omega_{C/T}$ at a fixed cavity detuning, $\delta_{C-X}=-15.4$~meV. 
    }
    \label{fig:suppl:EvsRt}
\end{figure}
\begin{figure}
    \centering
    \includegraphics[width=0.6\columnwidth]{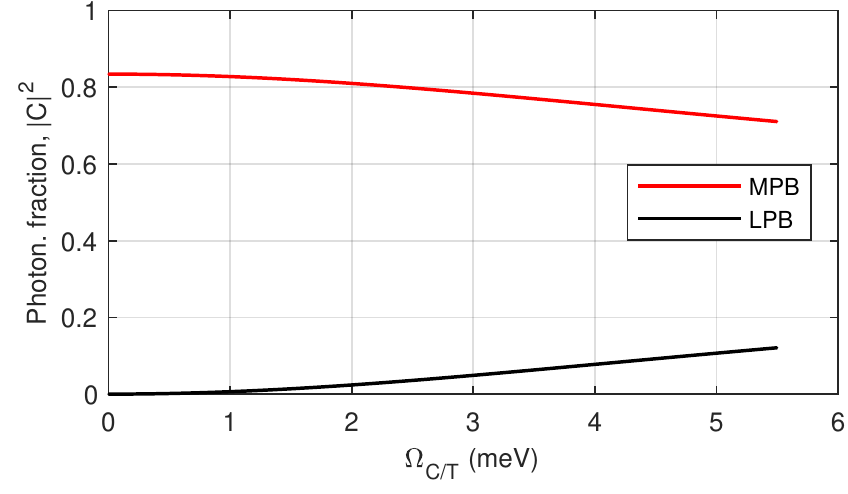}
    \caption{Photonic fraction, $\left|C\right|^2$, of the MPB and LPB as function of $\hbar\Omega_{C/T}$ at a fixed cavity detuning, $\delta_{C-X}=-15.4$~meV.}
    \label{fig:suppl:CvsRt}
\end{figure}

\subsection{Estimation of the total reservoir density in the case of nonlinear trion-polaritons}

The scattering of polaritons with the excitonic disorder potential (which we estimate in the order of $10$-$15$~meV from the inhomogeneous linewidths of exciton/trion emission) effectively populates the reservoir states~\cite{Sarkar2010,Snoke1997,Tollerud2016,Menard2014} with a scattering rate of the order of polariton linewidth $\gamma_{pol}\sim3-5$~meV. The polariton nonlinearity is driven by the total density of excitons and trions $n_{tot}$ excited in the system with a single pulse, i.e excitons and trions due to polariton density and due to reservoir. By measuring the total number of photons transmitted through the microcavity (MC) in a single pulse $n_{phot}$ we can estimate this density as
\begin{equation}
n_{tot} = n_{phot}\cdot\left(\left|X\right|^2+\left|T\right|^2+\frac{\gamma_{pol}}{\gamma_C\cdot\left|C\right|^2} \right).\label{eq:dens}
\end{equation}
Here $\left|X\right|^2$, $\left|T\right|^2$, $\left|C\right|^2$ are exciton, trion and photon fractions of polariton. The bare cavity mode linewidth $\gamma_C \sim 0.4$~meV corresponds to the photon lifetime of $\tau_C\simeq 4$~ps. The third term in the above formula is dominant. It describes the ratio of absorbed to radiatively escaped polaritons,  assuming that the lifetime of the reservoir is much longer than cavity photon lifetime and that there is no backscattering from the reservoir to polariton states. For the case of resonantly driven trion-polaritons the reservoir and hence $n_{tot}$ consists of mostly trions, since the peak of exciton density of states is blue detuned by several ten's of meV from the trion.

To estimate $n_{tot}$, one has to measure the number of photons $n_{phot}$ emitted by the cavity in a single pulse after the resonant excitation with the 100-fs pulsed laser. To do that we shifted the position of the photon mode to the very negative detuning about 10 nm below the trion level and ramped up the power of the laser so that the power of the light transmitted through the microcavity is about 50 nW. The corresponding photon counts of the transmitted light on the CCD were measured as a reference value. By relating the measured photon counts of light emitted by polariton microcavity  to this reference value it is possible to deduce the average power of the polariton emission in W, $I_{pol}$. The density of photons emitted by microcavity in a single pulse is then calculated as $n_{phot}=\frac{I_{pol}}{f\hbar\omega A}$, where $\hbar\omega$ is the polariton energy, $f=10^3$ Hz is the repetition rate of our pulsed laser and $A \approx 3$ $\mu$m$^2$ is the mode area.

\begin{figure}
    \centering
    \includegraphics[width=0.6\columnwidth]{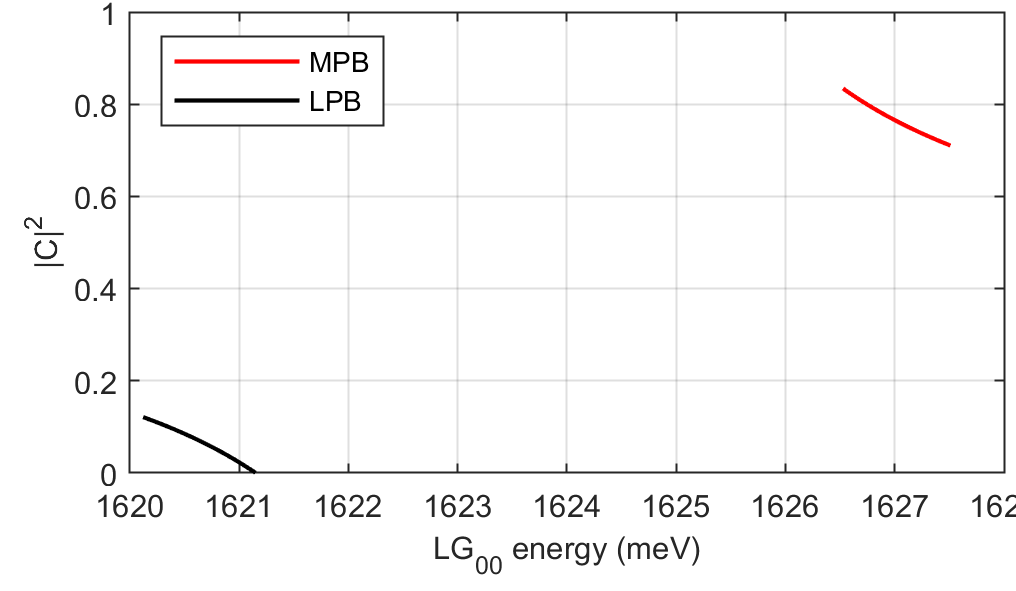}
    \caption{Combined $E_{\LGOO}$ vs $\left|C\right|^2$ calibration curves for MPB and LPB for a fixed cavity detuning, $\delta_{C-X}=-15.4$~meV.}
    \label{fig:suppl:Tcalib}
\end{figure}

The Hopfield coefficients used in Eq. \eqref{eq:dens} for the total reservoir density also have to be determined for each pump power, since the change of either cavity-trion or cavity-exciton Rabi-splitting would change these values. To illustrate this we fix the cavity-exciton Rabi-splitting and plot theoretical peak positions of LPB and MPB modes as a function of cavity-trion Rabi-splitting, which is shown in Fig.~\ref{fig:suppl:EvsRt} for the cavity mode-exciton detuning used in Fig.~3 [main text]. The theoretical dependence of the photonic fraction of the polariton branches on the cavity-trion Rabi-splitting is shown in Fig.~\ref{fig:suppl:CvsRt}. Combining these two sets of data one can produce the ``calibration'' dependencies of the Hopfield coefficients. The photonic fractions $|C|^2$ vs MPB and LPB energies are shown in Fig.~\ref{fig:suppl:Tcalib}. This can be used to infer the values of the actual Hopfield coefficients of MPB branch in Fig.~3 of the main text for each pump power.

This method is valid since in the experiment cavity-trion Rabi-splitting is completely quenched before any effect is seen on the cavity-exciton Rabi-splitting, as discussed in the main text.  Thus, to estimate total reservoir density we, first, determined the peak position of MPB mode by fitting each measured spectra, and then using this value we determined effective photon, exciton, and trion fractions of the MPB mode from the calibration curves. Finally, the total reservoir density for each pump power was calculated using Eq. \eqref{eq:dens}.

\subsection{Estimation of the total reservoir density in the case of nonlinear neutral exciton-polaritons }

\begin{figure}
\centering
\includegraphics[width=0.6\columnwidth]{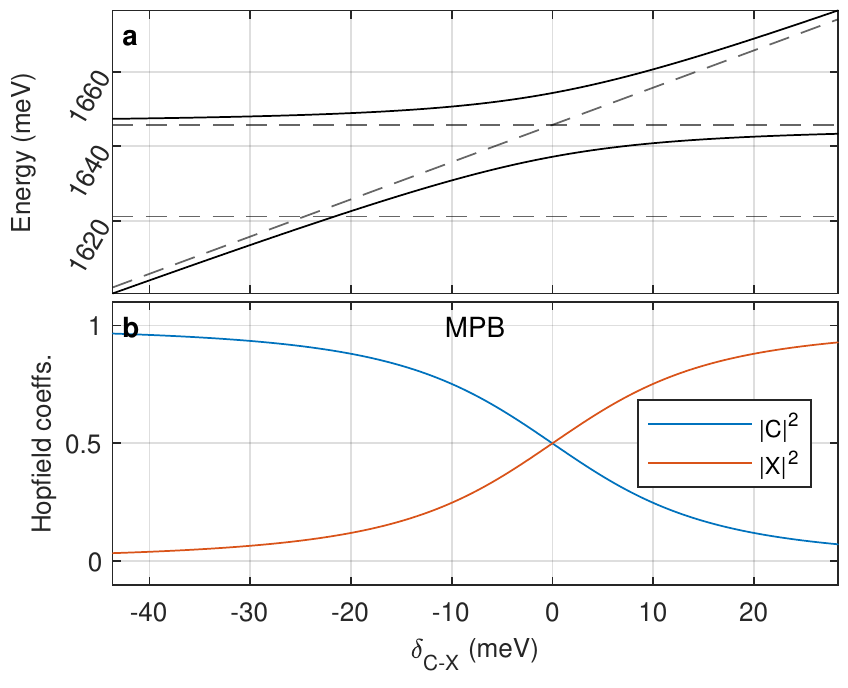}
\caption{\textbf{a} Polariton dispersions (solid lines) and bare cavity, exciton, and trion (dashed lines) from Fig.~1d [main text] assuming no coupling with trions. \textbf{b}  $\left|C\right|^2, \left|X\right|^2$ Hopfield coefficients of the middle polariton branch.}
\label{fig:suppl:Hopf2}
\end{figure}

Since the neutral exciton-polariton nonlinearities are weak in comparison to trion-polaritons they are studied at high pump when the strong coupling between photons and trions is quenched. In this case the polariton system can now be described by a simpler two-oscillator model,
\begin{equation}
    \begin{bmatrix}
    E_C(L) & \frac{1}{2}\hbar\Omega_{X} \\
    \frac{1}{2}\hbar\Omega_{X} & E_X 
    \end{bmatrix} \psi = E \psi, \label{eq:model}
\end{equation}
which results in two polariton branches UPB and LPB. $\psi$ is a two-component basis vector. Separate MPB and LPB do not longer exist, since they recombine into a single MPB. This model is applicable for all polariton densities $>200~\iums$, which is the case for all data shown in Fig.~4 [main text] and portion (above $200~\iums$) in Fig.~3 [main text]. The polariton mode positions and the corresponding Hopfield coefficients for the MPB are shown in Fig.~\ref{fig:suppl:Hopf2}. Table~\ref{tbl:hopfCXonly} shows the values of the Hopfield coefficients for the same cavity mode-exciton detunings as in Table~\ref{tbl:hopf} but now assuming that there is no trion-cavity mode coupling.

\begin{table}[]
\begin{tabular}{l|l|l|l|l|l|l|l|l|}
\hline
\multicolumn{1}{|l|}{$\delta_{C-X}=$} & \multicolumn{2}{l|}{$-15.4$ meV} & \multicolumn{2}{l|}{$-2.4$ meV} & \multicolumn{2}{l|}{$+2.0$ meV} & \multicolumn{2}{l|}{$+8.8$ meV} \\ \hline
 & UPB & MPB & UPB & MPB & UPB & MPB & UPB & MPB \\ \hline
\multicolumn{1}{|l|}{$\left|C\right|^2$} & 0.166 & 0.834 & 0.430 & 0.570 & 0.557 & 0.443 & 0.729 & 0.271 \\ \hline
\multicolumn{1}{|l|}{$\left|X\right|^2$} & 0.834 & 0.166 & 0.570 & 0.430 & 0.443 & 0.557 & 0.271 & 0.729 \\ \hline
\end{tabular}
\caption{Fitted values of the Hopfield coefficients for UPB and MPB branches for the four experimental values of cavity-exciton detunings assuming $\Omega_{T}=0$.}
\label{tbl:hopfCXonly}
\end{table}

\begin{figure}
    \centering
    \includegraphics[width=0.6\columnwidth]{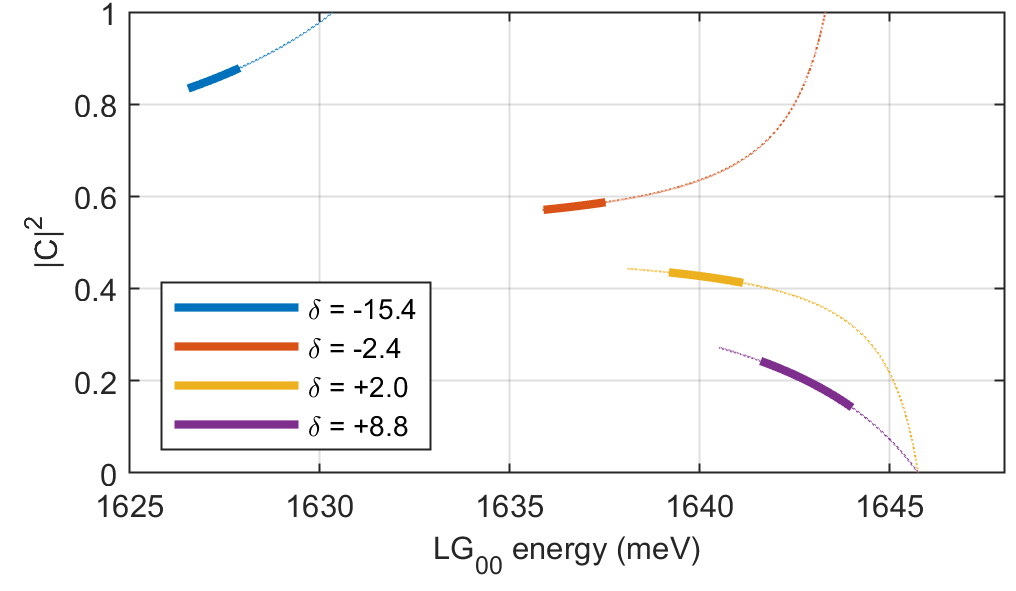}
    \caption{Calibration curves $E_{\LGOO}$ vs $\left|C\right|^2$ for the MPB with a two coupled oscillator model, obtained by varying the cavity-exciton Rabi-splitting from 0 to 19~meV for the four experimental cavity-exciton detunings. Bold solid sections of the curves correspond to the experimentally observed ranges of MPB peak positions for the corresponding detunings.}
    \label{fig:suppl:Xcalib}
\end{figure}

Using the same approach as discussed above, one can obtain the calibration curves ($E_{\mathrm{MPB}}$ vs $\left|C\right|^2$) for different cavity mode-exciton detunings by using this two-oscillator model and varying cavity-exciton Rabi-splitting. The data for the detunings used in the experiment are summarised in Fig.~\ref{fig:suppl:Xcalib}. Finally, the total exciton density for each exciton-photon detuning is deduced using formula \eqref{eq:dens} from section A.2 (there we assume that trion fraction $T=0$, since at high excitation density the trion-photon strong coupling is quenched).

\subsection{The strength of trion-polariton nonlinearity }

For a given energy of MPB the value of the Rabi-splitting $\Omega_T$ at each power density can be deduced using the calibration dependencies in Fig.~2.
The effective strength of trion nonlinearity responsible for the quenching of strong coupling is defined as a rate of decrease of the Rabi-splitting $\hbar \Omega_T$ with $n_{tot}$\cite{Brichkin2011}:
\begin{align}
\beta_T^{\mathrm{eff}}= -\delta \hbar \Omega_T/\delta n_{tot}.
\end{align}

Alternatively, in the first order approximation $\beta_T^{\mathrm{eff}}$ can related directly to the redshift of the MPB branch in Fig.~3b of the main text as follows
\begin{align}
\label{eq:beta_T_in_SM}
\beta_T^{\mathrm{eff}} = -\eta_T \delta E_{\mathrm{MPB}}/\delta n_{tot},
\end{align}
where $\eta_T = \sqrt{\delta^2_{\mathrm{P-T}}+(\hbar\Omega_{T})^2} / (\hbar\Omega_{T}/2)$ and $\delta_{\mathrm{P-T}}=+5.3$~meV is the energy detuning between the trion energy level and the lower polariton branch arising from coupling between photon and neutral exciton only (see above) for $n_{tot}<2\cdot10^2~\iums$. In Eq.~\eqref{eq:beta_T_in_SM} the minus sign explicitly accounts that Rabi frequency is decreasing function of density.

\begin{figure}
    \centering
    \includegraphics[width=0.6\columnwidth]{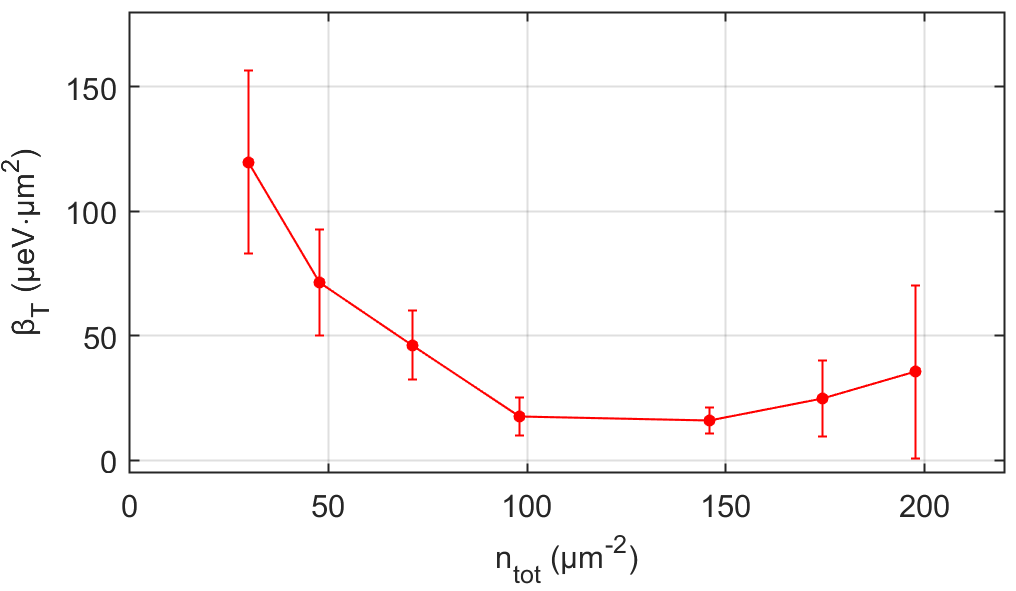}
    \caption{$\beta_T^{\mathrm{eff}}$ as a function of $n_{tot}$. The error bars are 95\% CI deduced taking into account the random error in the determination of the trion-polariton peak position at each power from the fitting procedure. }
    \label{fig:suppl:beta}
\end{figure}

In the first order approximation our theory predicts a constant value of $\beta_T^{\mathrm{eff}}$ with density, or, in other words, a linear reduction of the trion-polariton Rabi-splitting with density (see SM, sec.C). As it is seen in Fig.~3c of the main text there is a good qualitative agreement between the dependence of the trion Rabi splitting and the theoretical prediction as a function of trion density: the experimental average value of $\beta_T^{\mathrm{eff}}=37 \pm 3$ $\mu$eV$\mu$m$^2$ is in quantitative agreement with the theoretical estimate of $30$ $\mu$eV$\mu$m$^2$.
Nevertheless, it is seen that the experimental points in Fig.~3c of the main text are not precisely positioned on the straight theoretical line. This is reflected by the fact that the experimental values of $\beta_T^{\mathrm{eff}}$ (deduced from the experimental data in Fig.~3c of the main text) vary from $120$ to $20$ $\mu$eV$\cdot \mu$m$^2$ over the density range $n_{tot}$ from 0 to  200 $\mu$m$^{-2}$, which is shown in Fig.~\ref{fig:suppl:beta}. We believe this variation of $\beta_T^{\mathrm{eff}}$ with $n_{tot}$ observed in the experiment may arise from the higher order effects due to composite nature of trions, not included in the theoretical treatment (see SM, sec.C).

\subsection{Parameters characterising neutral exciton-polariton nonlinearity }

The energy blueshift of the neutral exciton polariton arises from two mechanism: (1) the reduction of exciton-photon Rabi-splitting $\hbar \Omega_X$, and/or (2) blueshift of the neutral exciton level $E_X$ due to Coulomb exchange interactions. 

Mechanism (1) is characterized by the rate of reduction of Rabi-splitting $\hbar \Omega_X$ with exciton density\cite{Brichkin2011}:   
\begin{align}
\beta_X^{\mathrm{eff}}=- \delta \hbar \Omega_X/\delta n_{tot}.
\end{align}
It can be deduced from the MPB polariton energy blueshift assuming that only mechanism (1) is dominant:
\begin{align}
\beta_X^{\mathrm{eff}} = \eta_X \delta E_{\mathrm{MPB}}/\delta n_{tot},
\end{align}
where $\eta_X = \sqrt{\delta^2_{\mathrm{C-X}}+(\hbar\Omega_{X})^2} / (\hbar\Omega_{X}/2)$ and $\delta_{\mathrm{C-X}}$ is the detuning between bare photon mode and exciton level for a given data set.

Mechanism (2) is characterised by the rate of the blueshift of the exciton level with exciton density~\cite{Brichkin2011}:
\begin{align}
g_X^{\textrm{eff}} = \delta E_{X}/\delta n_{tot}.
\end{align}

Similarly, assuming that only mechanism (2) contributes to polariton blueshift $g_X^{\textrm{eff}}$ can be related to the polariton energy shift~\cite{Brichkin2011}:
\begin{align}
g_X^{\textrm{eff}} = \xi_X \delta E_{\mathrm{MPB}}/\delta n_{tot},   
\end{align}
where $\xi_X= [1/2 + \delta_{C-X}/(2\sqrt{\delta^2_{C-X}+(\hbar\Omega_{X})^2})]^{-1}$ is the inverse of an excitonic fraction.

In the experiment we can measure only the nonlinear behaviour (blueshift) of MPB. The UPB cannot be measured due to tunability of our laser. Therefore, experimentally we cannot separate the contributions to the neutral exciton-polariton optical nonlinearity from mechanisms (1) and (2). However, assuming that either only mechanism (1) or (2) is responsible for the blueshift of MPB, we can deduce the dependencies of \textit{the upper limits} of $\beta_X^{\textrm{eff}} = -d (\hbar \Omega_X) / d n_{tot}$ and  $g_X^{\textrm{eff}} = d E_X / d n_{tot}$ factors on exciton density~\cite{Brichkin2011}, respectively. In the main text (Fig.~5) we show that the theoretical $g_X^{\textrm{th}}$ parameter is in semi-quantitative agreement with the experimental values of the upper limit of $g_X^{\textrm{eff}}$ in the range of exciton densities $3\cdot 10^3<n_{tot}<3\cdot 10^4 \;\mu$m$^{-2}$.

\begin{figure}
\centering
\includegraphics[width=0.6\columnwidth]{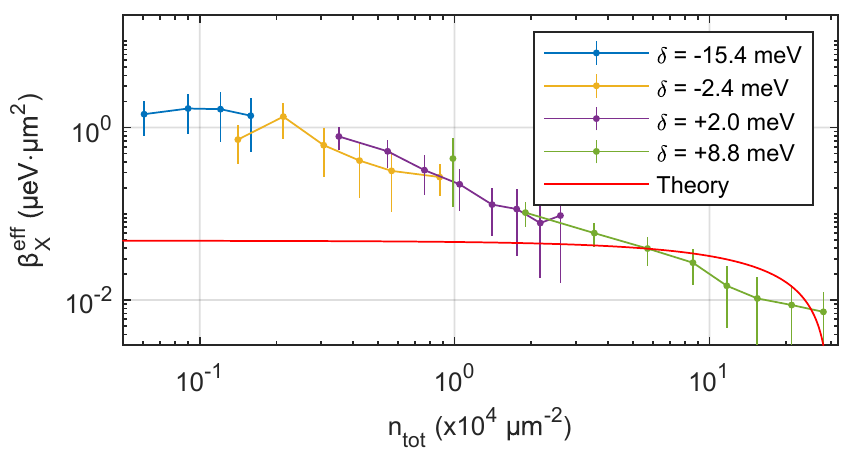}
\caption{The experimental upper limit of the effective interaction constant $\beta_X^{\textrm{eff}}$ as a function of the estimated exciton density, $n_{tot}$. The data correspond to four different cavity-exciton detunings ($\delta$): $+8.8$~meV (olive), $+2.0$~meV (purple), $-2.4$~meV (orange), and $-15.4$~meV (blue). The error bars ($95\%$ CI) are deduced taking into account errors in determining the MPB peak positions at each pump power (exciton density). The red solid curve corresponds to the theoretical values.}
\label{fig:suppl:betaX}
\end{figure}

Now let us assume instead that the mechanism (1) is the only dominant mechanism over the whole density range. In this case we can observe that  the theoretical $\beta_X^{\textrm{th}}$-factor is well below the experimental values of the upper limit of $\beta_X^{\textrm{eff}}$ at $n_{tot}< 10^4 \;\mu$m$^{-2}$ as shown in Fig.~\ref{fig:suppl:betaX}. $\beta_X^{\textrm{th}}$ approaches the experimental values only at higher densities $n_{tot}> 3\cdot10^4 \;\mu$m$^{-2}$. Such a discrepancy between the experiment and theory indicates that our assumption is incorrect; phase space filling for neutral exciton-polaritons [mechanism (1)] becomes important only at very high exciton densities $n_{tot}> 3\cdot10^4 \;\mu$m$^{-2}$, when the average distance between excited excitons is less than $5-6$ nm and becomes comparable to the exciton Bohr radius $a_B\sim1$ nm. By contrast, mechanism (2) is the dominant mechanism at intermediate exciton densities $n_{tot}<3\cdot 10^4 \;\mu$m$^{-2}$.

\subsection{Measurement of the cavity mode decay rate (photon lifetime) }

\begin{figure}
\centering
\includegraphics[width=0.6\columnwidth]{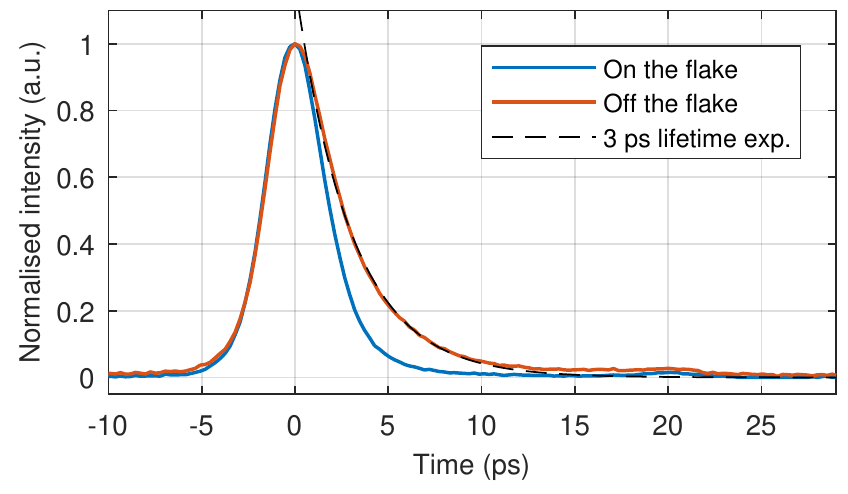}
\caption{Microcavity lifetime measured using a streak camera. Blue response measured with resonant transmission of a 100-fs pump pulse, when the cavity was on the flake. Orange curve corresponds to the bare cavity, off the flake, with no active region. Note that the detected signal profile is limited by the streak camera resolution of 1-2 ps and possible laser pulse jitter.}
\label{fig:suppl:lifetime}
\end{figure}

The bare cavity mode decay rate ($\gamma_C$) was deduced from the temporal decay of emission intensity of the bare $\LGOO$ mode (without flake in the cavity) excited resonantly with 100-fs pulse. The measurements were performed using streak-camera with the resolution time of 2 ps (see Fig.~\ref{fig:suppl:lifetime}). We obtained the experimental cavity lifetime of approximately 4~ps, which corresponds to the FWHM of bare cavity mode $\sim 400~\mu$eV, the value we measured on the spectrometer.

\subsection{Nonlinear refractive index $n_2$ due to trion-polaritons. Comparison with other systems. }

\subsubsection{Nonlinear refractive index $n_2$ of hybrid microcavity-MoSe$_2$ polariton system. }

We note that applications of 2D materials imply that they would be integrated into photonic structures made of bulk semiconductors/dielectrics. Therefore, when characterising nonlinear optical properties of 2D materials it is useful to consider the nonlinearity of the whole hybrid 2D materials-semiconductor/dielectric photonic system and compare it with that of bulk photonic materials.

In order to compare trion-polariton nonlinearity with Kerr-like optical nonlinearity observed in bulk materials we can treat our open-access microcavity system with embedded MoSe$_2$ as a microcavity filled with a bulk of some nonlinear optical material, characterised by the effective nonlinear refractive coefficient $n_{2(MC)}$. The
$n_{2(MC)}$ coefficient due to trion-polariton nonlinearity can be estimated taking into account the effective optical path covered by a photon during the round trip between the two mirrors of the microcavity, which must be equal to an an integer number of polariton wavelengths,
\begin{align}
2(n_{eff}+n_{ 2(MC)}\frac{N_{pol}hc^2}{L\lambda})L=m(\lambda+\delta\lambda).
\end{align}
Here $n_{eff}$ is the effective refractive index of the microcavity, $N_{pol}$ is the total number of polaritons excited inside the MC with a single pulse, $L$ is the effective cavity length, $\lambda$ is the wavelength of the trion-polariton emission in free space, $\delta\lambda$ is the nonlinear shift of trion-polariton resonance, $m$ is the order of the longitudinal cavity mode coupled with the trion. $2n_{2(MC)}\frac{N_{pol}hc^2}{L\lambda}L$ is the nonlinear optical path acquired by photon during the round trip between the two mirrors. Given absorption is the dominant process in our system $N_{pol} \approx n_{tot}$.

Using Eqs. (5) and (10) we get the following expression
\begin{align}
n_{2(MC)}=\frac{n_{eff} L\beta_T^{\mathrm{eff}}\lambda^2}{h^2c^3\eta_T}
\end{align}
where $\eta_T = \sqrt{\delta^2_{\mathrm{P-T}}+(\hbar\Omega_{T})^2} / (\hbar\Omega_{T}/2)$, and $\delta_{\mathrm{P-T}}$ is the energy detuning between the trion energy level and the lower polariton branch arising from coupling between photon and neutral exciton only (see above). Taking into account the effective cavity refractive index $n_{eff}\approx1$ (since most of the cavity electromagentic field is confined in the gap between the two mirror), the wavelength of trion-polariton resonance $\lambda \approx 760$ nm, the effective cavity size $L\approx1$ $\mu$m, $\beta_T^{\mathrm{eff}}=37$ $\mu$eV$\mu$m$^2$ and $\eta_T=2$ ($\delta_{\mathrm{P-T}}=0$) we estimate  $n_{2(MC)}\sim1.4\cdot 10^{-13}$ $m^2$/W. This $n_{2(MC)}$ is about four to five orders of magnitude larger than $1.82\cdot 10^{-17}$ m$^2$/W in planar AlGaAs waveguides in the weak coupling regime~\cite{Belanger1997} and $6\times 10^{-18}$ m$^2$/W in silicon~\cite{Blanco-Redondo2014} and InGaP~\cite{Colman2010}, which have been used in a suspended membrane photonic crystal geometry.  Kerr nonlinear effects (optical bistability) have been investigated in slab photonic crystal Si microcavities with embedded graphene layer~\cite{Gu2012}. The effective $n_2$ of hybrid graphene-Si microcavity system has been derived to be of the order $n_2 \approx 7.7\cdot 10^{-17}$ m$^2$/W, which is $\sim3-4$ orders of magnitude less than $n_{2(MC)}$ due to trion-polariton nonlinearity. Finally, we note that the value $n_{2(MC)}\sim1.4\cdot 10^{-13}$ $m^2$/W due to trion polariton nonlinearity is an order of magnitude higher than $n_{2}\sim1\cdot 10^{-14}$ $m^2$/W reported by us in neutral exciton-polariton GaAs-based system~\cite{Walker2015}.

\subsubsection{Effective nonlinear refractive index $n_{2(MoSe_2)}$ per single TMDC monolayer arising from trion-polariton nonlinearity.}

To the best of our knowledge no Kerr-like nonlinear optical effects were studied in microcavities with embedded TMDC materials in the weak light-matter coupling regime. However, there were several studies of the effects associated with Kerr-like optical nonlinearity of bare layers of TMDCs (MoSe$_2$, MoS$_2$, MoTe$_2$) and graphene in the weak light-matter coupling regime on a picosecond timescale~\cite{Wang2014}. The values of $n_2$ coefficients for TMDCs layers were measured in the range $10^{-16}$-$10^{-17}$ m$^2$/W depending on the excitation energy (above or below band gap).  
The reference [\onlinecite{Torres-Torres2016}] reports the $n_2$ coefficient for WS$_2$ monolayer to be about $1.1 \cdot 10^{-15}$ $m^2$/W.
The value of $n_2$ coefficient for pure graphene flakes was measured about $2\cdot 10^{-15}$ m$^2$/W~\cite{Wang2014}.
The reference [\onlinecite{Gu2012}], which studied nonlinear hybrid Si-graphene microcavity, deduces $n_2$ coefficient for a single  graphene to be of the order $10^{-13}$ m$^2$/W.

In our trion-polariton microcavity system we can derive the effective nonlinear refractive index $n_{2(MoSe_2)}$ per single TMDC monolayer taking into account that in reality the nonlinear optical phase is acquired by light only on the passage of the monolayer during the round trip between the mirrors. Therefore, $n_{2(MoSe_2)}$ can be simply obtained by normalising $n_{2(MC)}$ to $d_{MoSe_2}/L$, where $d_{MoSe_2}\sim 1$ nm is the MoSe$_2$ thickness, yielding $n_{2(MoSe_2)}\sim 1.4\cdot 10^{-10}$ $m^2$/W. This value is at least five (three) orders of magnitude larger than in TMDC 2D materials (graphene) studied in the weak light-matter coupling regime without formation of polaritons. 

\section{Nonlinear neutral exciton-polaritons: theory}

In the first section we describe the exciton-photon coupled system, accounting for the composite electron-hole (e-h) nature of neutral exciton. The case of a trion mode coupled to the optical mode is considered in the next section.

\subsection{Exciton-polariton Rabi-splitting} To start, we consider an optical cavity described by the bosonic annihilation and creation operators $\hat{c}$ and $\hat{c}^\dagger$, such that their commutation relations are $[\hat{c},\hat{c}^\dagger]=1$, $[\hat{c}^\dagger,\hat{c}^\dagger]=[\hat{c},\hat{c}]=0$. Coupled to a semiconducting medium, an optical photon creates an exciton, corresponding to the bound electron-hole pair. The creation of an electron with the momentum $\mathbf{q}$ is described by the fermionic operator $\hat{a}_{\mathbf{q}}^\dagger$, and hole creation is described by the operator  $\hat{b}_{\mathbf{q}}^\dagger$. Their corresponding anti-commutation relations read $\{ \hat{a}_{\mathbf{q}}, \hat{a}_{\mathbf{q}'}^\dagger \} = \hat{a}_{\mathbf{q}} \hat{a}_{\mathbf{q}'}^\dagger + \hat{a}_{\mathbf{q}'}^\dagger \hat{a}_{\mathbf{q}} = \delta_{\mathbf{q},\mathbf{q}'}$, $\{ \hat{a}_{\mathbf{q}}, \hat{a}_{\mathbf{q}'} \} = \{ \hat{a}_{\mathbf{q}}^\dagger, \hat{a}_{\mathbf{q}'}^\dagger \} = 0$, where $\delta_{\mathbf{q},\mathbf{q}'}$ is Kronecker delta function (and the same holds for $\hat{b}_{\mathbf{q}}$). Accounting for the attractive Coulomb interaction between an electron and a hole, the excitonic operator can be written as a composite boson $\hat{X}_{\nu}$, where $\nu$ is a general index which denotes the center-of-mass (CM) and internal degrees of freedom. The transition between electron-hole and composite exciton picture follows as $\hat{X}_{\nu}^\dagger = \sum_{\mathbf{k}_\alpha, \mathbf{k}_\beta} \langle \mathbf{k}_\beta, \mathbf{k}_\alpha| \nu \rangle \hat{a}_{\mathbf{k}_\alpha}^\dagger \hat{b}_{\mathbf{k}_\beta}^\dagger$. Here, $\mathbf{k}_{\alpha,\beta}$ correspond to electron and hole momenta, and $\langle \mathbf{k}_\beta, \mathbf{k}_\alpha| \nu \rangle$ is an exciton wave function written in the momentum space. The reversed transformation for describing an electron-hole pair in bosonic language can be written as $\hat{a}_{\mathbf{k}_\alpha}^\dagger \hat{b}_{\mathbf{k}_\beta}^\dagger  = \sum_{\nu} \langle \nu|  \mathbf{k}_\beta, \mathbf{k}_\alpha \rangle \hat{X}_{\nu}^\dagger$, where summation goes over possible states of composite excitons $\nu$ in the appropriate orthonormal basis, such that $\sum_\nu |\nu \rangle \langle \nu| = \mathbf{1}$. 

In the following we are interested in the system with strong light-matter coupling, where a composite exciton of certain CM momentum is coupled to the cavity mode. The Hamiltonian for the considered system reads
\begin{align}
\label{eq:H_tot}
\hat{\mathcal{H}} = \hat{H}_{\mathrm{cav}} + \hat{H}_X + \hat{H}_{\mathrm{coupl}},
\end{align}
where the first and second terms describe the free energy for the cavity photon mode, $\hat{H}_{\mathrm{cav}} = \sum_{\mathbf{q}} \omega_{\mathrm{cav},\mathbf{q}} \hat{c}_{\mathbf{q}}^\dagger \hat{c}_{\mathbf{q}}$ ($\hbar = 1$ hereafter), and composite exciton $\hat{H}_X$ (i.e. coupled electron-hole) Hamiltonians. $\omega_{\mathrm{cav},\mathbf{q}} $ denotes the two-dimensional dispersion for the planar cavity mode, with typically ultralow mass, such that only small $q$'s are considered. The third term describes the coupling between light and matter excitations. It can be written as a creation of an electron-hole pair by the cavity field with a coupling constant $g$,
\begin{align}
\label{eq:H_coupl}
\hat{H}_{\mathrm{coupl}} = \sum_{\mathbf{k}_\alpha,\mathbf{k}_\beta,\mathbf{q}} (g \hat{a}_{\mathbf{k}_\alpha + \mathbf{q}}^\dagger \hat{b}_{\mathbf{k}_\beta}^\dagger \hat{c}_{\mathbf{q}} + h.c.) = \sum_{\mathbf{k}_\alpha,\mathbf{k}_\beta,\mathbf{q}} \sum_{i}  (g \langle i|  \mathbf{k}_\beta, \mathbf{k}_\alpha \rangle \hat{X}_{i}^\dagger \hat{c} + h.c.),
\end{align}
where in the second equality we have exploited excitonic form for the electron-hole pair, and considered dipolar transition with negligible transferred cavity momentum, $\hat{c}_{\mathbf{q}\rightarrow 0} \equiv \hat{c}$, being a usual assumption for description of strong coupling. Here, $g = \frac{e p_{cv}}{m} \sqrt{\frac{\hbar^2}{ 2\epsilon \epsilon_0 \omega_{\mathrm{cav}} L_{\mathrm{cav}} A}}$, where $p_{cv}$ is a matrix element for valence-to-conduction band transition, $m$ is a free electron mass, $L_{\mathrm{cav}}$ is a cavity length, $A$ is an area of the system. We consider an exciton mode at the fixed center-of-mass momentum, which for brevity is set to zero, $\hat{X}_0$, and derive the corresponding Heisenberg equations of motion. It reads
\begin{align}
\label{eq:dXdt}
i\frac{d \hat{X}_0}{d t} = [\hat{X}_0, \hat{H}_X] + [\hat{X}_0, \hat{H}_{\mathrm{coupl}}] = [\hat{X}_0, \hat{H}_X] + g \sum_{\mathbf{k}_\alpha,\mathbf{k}_\beta} \sum_{i}  \langle i|  \mathbf{k}_\beta, \mathbf{k}_\alpha \rangle \Big[\hat{X}_0, \hat{X}_{i}^\dagger \Big] \hat{c}.
\end{align}
The first term generally describes the energy $\omega_X$ at which the excitonic mode oscillates. The second, being proportional to $\hat{c}$ operator, provides the coupling to photonic mode, which we generally denote as $G = g \sum_{\mathbf{k}_\alpha,\mathbf{k}_\beta} \sum_{i}  \langle i|  \mathbf{k}_\beta, \mathbf{k}_\alpha \rangle \Big[\hat{X}_0, \hat{X}_{i}^\dagger \Big]$. If exciton corresponds to an ideal boson, i.e. $[\hat{X},\hat{X}^\dagger] = 1$, the coupling term reduces to $G = g \sum_{\mathbf{k}_\alpha,\mathbf{k}_\beta} \langle i|  \mathbf{k}_\beta, \mathbf{k}_\alpha \rangle \equiv g \sum_{\mathbf{k}} \phi_{\mathbf{k}}^* =: \Omega_X^{(0)}/2$, where we introduced the relative electron-hole momentum $\mathbf{k}$ and the Fourier transform of the exciton wave function $\phi_{\mathbf{k}}$. This energy corresponds directly to the Rabi energy for the light-matter coupled system. Performing the diagonalization of the system at zero detuning ($\omega_{\mathrm{cav}} = \omega_X$), the Rabi-splitting between normal modes of the system is equal to $\Omega_X^{(0)}$.

We proceed by considering the composite structure of exciton, which is formed by two fermions. This comes from the fact that creation of a (correlated) electron-hole pair is not equivalent to a boson creation, as long as number of created pairs grows. It originates from the Pauli exclusion principle, which does not allow certain pair configurations in the full fermionic treatment, while disregarded in the purely bosonic picture \cite{CombescotReview,KiraKoch}. The details for the difference between two cases were worked out by Monique Combescot and co-workers, and summarized in the so-called \textit{coboson approach} to excitonic systems \cite{CombescotReview}. In the following, we apply use the coboson formalism to find corrections to Rabi and exciton energy appearing due to effects of non-bosonicity.

The main consequence of the composite nature of exciton is its peculiar statistics, which resembles bosonic one for small e-h pair concentration $n \equiv N/A$, but changes once it becomes comparable to the inverse of the effective exciton area. The generic commutation relations between composite bosons can be formulated as (see Ref. [\onlinecite{CombescotReview}], eq. [4.16])
\begin{align}
\label{eq:comm_X}
[\hat{X}_m, \hat{X}_i^\dagger] \equiv \delta_{m,i} - \hat{D}_{mi},
\end{align}
where an operator $\hat{D}_{mi}$ describes the deviation from bosonicity for excitons due to its composite nature.

In particular, this can be observed when one writes the commutator in Eq. \eqref{eq:comm_X} using the expression for composite exciton with zero CM momentum and relative momentum $\mathbf{k}$, which is described by $\hat{X}_0^\dagger = \sum_{\mathbf{k}} \langle \mathbf{k} |0\rangle \hat{a}_{\mathbf{k}} \hat{b}_{-\mathbf{k}} \equiv \sum_\mathbf{k} \phi_{\mathbf{k}} \hat{a}_{\mathbf{k}} \hat{b}_{-\mathbf{k}}$. The commutator reads
\begin{align}
\notag
&[\hat{X}_0, \hat{X}^\dagger_0] = \sum_{\mathbf{k}_1} \sum_{\mathbf{k}_2} (\phi_{\mathbf{k}_1} \phi_{\mathbf{k}_2}^* \hat{a}_{\mathbf{k}_1} \hat{b}_{-\mathbf{k}_1} \hat{b}_{-\mathbf{k}_2}^\dagger \hat{a}_{\mathbf{k}_2}^\dagger  - \phi^*_{\mathbf{k}_2} \phi_{\mathbf{k}_1}  \hat{b}_{-\mathbf{k}_2}^\dagger \hat{a}_{\mathbf{k}_2}^\dagger \hat{a}_{\mathbf{k}_1} \hat{b}_{-\mathbf{k}_1}) = \sum_{\mathbf{k}_1} \sum_{\mathbf{k}_2} \phi_{\mathbf{k}_1} \phi_{\mathbf{k}_2}^* (\delta_{\mathbf{k}_1,\mathbf{k}_2} - \hat{a}_{\mathbf{k}_1}^\dagger \hat{a}_{\mathbf{k}_2} \delta_{\mathbf{k}_1,\mathbf{k}_2} \\ 
& - \hat{b}_{-\mathbf{k}_1}^\dagger \hat{b}_{-\mathbf{k}_2} \delta_{\mathbf{k}_1,\mathbf{k}_2})  = 1 - \sum_{\mathbf{k}} |\phi_{\mathbf{k}}|^2 (\hat{a}_{\mathbf{k}}^\dagger \hat{a}_{\mathbf{k}} + \hat{b}_{\mathbf{k}}^\dagger \hat{b}_{\mathbf{k}}).
\label{eq:comm}
\end{align}
The explicit form for the deviation operator is $\hat{D}_{00} =  \sum_{\mathbf{k}} |\phi_{\mathbf{k}}|^2 (\hat{a}_{\mathbf{k}}^\dagger \hat{a}_{\mathbf{k}} + \hat{b}_{\mathbf{k}}^\dagger \hat{b}_{\mathbf{k}})$. Its structure thus hints that the deviation depends on the electron (or exciton) number $N$.

To calculate the influence of the non-bosonicity on Rabi energy renormalization we need to estimate the expectation value of the last term in Eq. \eqref{eq:dXdt} considering the (unnormalized) many-coboson state $|N \rangle = (\hat{X}_0^\dagger)^N |\text{\o} \rangle$, and singling out the prefactor in front of the cavity photon operator $\hat{c}$. Here, $|\text{\o} \rangle$ denotes coboson vacuum state, and corresponding norm reads $\sqrt{\langle N | N \rangle} =\langle \text{\o} | \hat{X}_0^N (\hat{X}_0^\dagger)^N |\text{\o} \rangle^{1/2}$. Note that in the case of composite bosons it was shown to differ exponentially from an ideal boson normalization for large $N$,\cite{CombescotReview} though for physical observables the difference appears as higher order terms in small $N$ expansion. 

The expectation value for the commutator can be written as 
\begin{align}
\label{eq:Omega_N}
G(N) := \frac{ \langle N| g \sum_{\mathbf{k}_\alpha,\mathbf{k}_\beta} \sum_{i}  \langle i|  \mathbf{k}_\beta, \mathbf{k}_\alpha \rangle \Big[\hat{X}_0, \hat{X}_{i}^\dagger \Big] | N\rangle }{\langle N| N \rangle}= g \frac{\langle N| \sum_{\mathbf{k}} \phi_{\mathbf{k}}^* | N\rangle}{\langle N|N\rangle} - g \frac{\langle N| \sum_{\mathbf{k}_\alpha,\mathbf{k}_\beta} \sum_{i}  \langle i|  \mathbf{k}_\beta, \mathbf{k}_\alpha \rangle \hat{D}_{0i} | N\rangle}{\langle N|N\rangle}.
\end{align}
The first term in Eq. \eqref{eq:Omega_N} yields $g\sum_{\mathbf{k}}\phi_{\mathbf{k}}^*$ and is simply a Rabi frequency in the dilute system limit. The second term, however, involves the non-bosonicity operator. Accounting for the many-coboson state explicitly, it can be rewritten as
\begin{align}
\label{eq:Omega_N_2}
 g \frac{\langle \text{\o} | \hat{X}_0^N \sum_{\mathbf{k}_{\alpha,\beta}} \sum_{i}  \langle i|  \mathbf{k}_\beta, \mathbf{k}_\alpha \rangle \hat{D}_{0i} (\hat{X}_0^\dagger)^N| \text{\o}\rangle}{\langle N|N\rangle} = g \frac{\langle \text{\o} |  \sum_{\mathbf{k}_{\alpha,\beta}} \sum_{i}  \langle i|  \mathbf{k}_\beta, \mathbf{k}_\alpha \rangle \hat{X}_0^N \Big[\hat{D}_{0i}, (\hat{X}_0^\dagger)^N \Big]| \text{\o}\rangle}{\langle N|N\rangle},
\end{align}
where we have accounted for the fact that the action of the deviation operator on the ground state gives 0, i.e. $\hat{D}_{0i}|\text{\o}\rangle = 0\cdot |\text{\o}\rangle$. Thus, its estimation relies on the commutator of the deviation operator with an exciton creation operator to the power $N$. For the first power, this can be derived as\cite{CombescotReview}
\begin{align}
\label{eq:comm_D_with_X}
\left[ \hat{D}_{mi}, \hat{X}_j^\dagger \right] = \sum_n \left[ \lambda \left( \begin{array}{cc}
    n & j \\
    m & i \end{array} \right) +
\lambda \left( \begin{array}{cc}
    m & j \\
    n & i \end{array} \right)
    \right] \hat{X}_n^\dagger,
\end{align}
where $\lambda$ denotes the Pauli scattering element for input indices $(i,j)$ and output $(n,m)$. It reads explicitly
\begin{align}
\label{eq:lambda_gen}
\lambda \left( \begin{array}{cc}
    n & j \\
    m & i \end{array} \right) = \int d\mathbf{r}_{\alpha_1} d\mathbf{r}_{\alpha_2} d\mathbf{r}_{\beta_1} d\mathbf{r}_{\beta_2} \phi_{m}^*(\mathbf{r}_{\alpha_1},\mathbf{r}_{\beta_2}) \phi_{n}^*(\mathbf{r}_{\alpha_2},\mathbf{r}_{\beta_1}) \phi_{i}(\mathbf{r}_{\alpha_1},\mathbf{r}_{\beta_1}) \phi_{j}(\mathbf{r}_{\alpha_2},\mathbf{r}_{\beta_2}),
\end{align}
where $\phi_{i}(\mathbf{r}_{\alpha_1},\mathbf{r}_{\beta_1})$ is a generic coboson wavefunction written in the real space representation. It can be rewritten as a product of CM and relative motion component, $\phi_{i}(\mathbf{r}_{\alpha},\mathbf{r}_{\beta}) = (e^{i\mathbf{Q}_i \cdot \mathbf{R}_{\alpha \beta}}/\sqrt{A}) \langle \mathbf{r}_{\alpha \beta}| i \rangle$, where $\mathbf{R}_{\alpha \beta}$ and $\mathbf{r}_{\alpha \beta}$ are CM and relative coordinates for e-h pair, which correspond to coboson with CM momentum $\mathbf{Q}_i$ and relative motion quantum number $i$. The relative motion wavefunction can be also rewritten in the momentum space as $\langle \mathbf{r}| i \rangle = \sum_{\mathbf{k}} \langle \mathbf{r}| \mathbf{k} \rangle \langle \mathbf{k} | i\rangle = \sum_{\mathbf{k}} (e^{i\mathbf{k} \cdot \mathbf{r}}/\sqrt{A}) \langle \mathbf{k}| i \rangle$, which we will use in future.

Proceeding with the estimation for the influence of the deviation in the many-coboson state, the commutator with $N$-exciton creation operator reads
\begin{align}
\label{eq:comm_D_with_X^n}
\left[ \hat{D}_{0i}, (\hat{X}_j^\dagger)^N \right] = N (\hat{X}_0^\dagger)^{N-1} \sum_n \left[ \lambda \left( \begin{array}{cc}
    n & i \\
    0 & 0 \end{array} \right) +
\lambda \left( \begin{array}{cc}
    0 & i \\
    n & 0 \end{array} \right)
    \right] \hat{X}_n^\dagger.
\end{align}
Using this, Eq. \eqref{eq:Omega_N} can be rewritten as
\begin{align}
\label{eq:Omega_N_res}
G(N) = g \sum_{\mathbf{k}} \phi_{\mathbf{k}}^* - g N \sum_{\mathbf{k}_{\alpha,\beta}} \sum_{i,n}  \langle i|  \mathbf{k}_\beta, \mathbf{k}_\alpha \rangle \left\{ \lambda \left( \begin{array}{cc}
    n & i \\
    0 & 0 \end{array} \right) +
\lambda \left( \begin{array}{cc}
    0 & i \\
    n & 0 \end{array} \right) \right\} \langle \text{\o} | \hat{X}_0^N \hat{X}_n^\dagger (\hat{X}_0^\dagger)^{N-1} | \text{\o}\rangle / \langle N|N\rangle.
\end{align}
One can immediately see in the second term on the RHS that the Rabi frequency depends on exciton concentration $N$, times the prefactors coming from Pauli scattering elements.

First, let us consider the intuitively easy case where the internal coboson index $i$ coincides with the mode of interest, labeled as $0$. Later, we show that this corresponds to the lower-order-in-$N$ correction. In this case the expectation value $\langle \text{\o} | \hat{X}_0^N \hat{X}_n^\dagger (\hat{X}_0^\dagger)^{N-1} | \text{\o}\rangle / \langle N|N\rangle |_{n = 0} = 1$. Considering the first Pauli scattering term $\lambda(0,i;0,0)$, the summation over internal coboson index is performed as
\begin{align}
&\sum_{\mathbf{k}} \sum_i \lambda \left( \begin{array}{cc}
    0 & i \\
    0 & 0 \end{array} \right) \langle i|  \mathbf{k} \rangle = \frac{1}{A^2} \sum_{\mathbf{k}} \int d\mathbf{r}_{\alpha_1} d\mathbf{r}_{\alpha_2} d\mathbf{r}_{\beta_1} d\mathbf{r}_{\beta_2} \langle 0 |\mathbf{r}_{\alpha_1}-\mathbf{r}_{\beta_2} \rangle \langle 0 |\mathbf{r}_{\alpha_2}-\mathbf{r}_{\beta_1}\rangle \langle \mathbf{r}_{\alpha_1}-\mathbf{r}_{\beta_1}|0\rangle \sum_i \langle  \mathbf{r}_{\alpha_2}-\mathbf{r}_{\beta_2} |i\rangle \langle i|  \mathbf{k} \rangle \notag \\
&= \sum_{\mathbf{k},\mathbf{k}_1,\mathbf{k}_2,\mathbf{k}_3} \frac{1}{A^4} \int d\mathbf{r}_{\alpha_1} d\mathbf{r}_{\alpha_2} d\mathbf{r}_{\beta_1} d\mathbf{r}_{\beta_2} e^{-i \mathbf{k}_1 \cdot (\mathbf{r}_{\alpha_1}-\mathbf{r}_{\beta_2})} e^{-i \mathbf{k}_2 \cdot (\mathbf{r}_{\alpha_2}-\mathbf{r}_{\beta_1})} e^{i \mathbf{k}_3 \cdot (\mathbf{r}_{\alpha_1}-\mathbf{r}_{\beta_1})} e^{i \mathbf{k} \cdot (\mathbf{r}_{\alpha_2}-\mathbf{r}_{\beta_2})} \times \notag \\
&\times \langle 0|\mathbf{k}_1 \rangle \langle 0|\mathbf{k}_2 \rangle \langle \mathbf{k}_3|0 \rangle = \sum_{\mathbf{k}} |\phi_{\mathbf{k}}|^2 \phi_{\mathbf{k}}.
\end{align}
Here, for passing through the first equation sign we used that: 1) coboson states form a full orthonormal basis, $\sum_i |i\rangle \langle i| = \mathbf{1}$; 2) transition element between between real and momentum space reads $\langle  \mathbf{r}_{\alpha_2}-\mathbf{r}_{\beta_2} |  \mathbf{k} \rangle = e^{i \mathbf{k} \cdot (\mathbf{r}_{\alpha_2}-\mathbf{r}_{\beta_2})}/\sqrt{A}$. For the second equality we exploited Dirac delta function definition in 2D, being $\int d\mathbf{r} e^{i\mathbf{k}\cdot \mathbf{r}} = A \delta_{\mathbf{k},\mathbf{0}}$, which reduces summation to a single index, and recall our definition $\langle \mathbf{k} |0\rangle \equiv \phi_{\mathbf{k}}$. The second Pauli scattering term gives the same contribution. 

In the case of 1s neutral exciton in a two-dimensional material the relation motion part of wavefunction in real space can be written as $\phi(\mathbf{r}) = \sqrt{2/\pi a_B^2} \exp(-r/a_B)$, with $\mathbf{r}$ being relative (e-h) coordinate. The momentum space version then reads
\begin{align}
\label{eq:phi_k}
\phi_{\mathbf{k}} = \sqrt{\frac{8\pi a_B^2}{A}} (1 + a_B^2 k^2)^{-3/2},
\end{align}
where $a_B$ corresponds to the 2D variational parameter.

Collecting everything together and performing summation as $\sum_{\mathbf{k}} \rightarrow \frac{A}{(2\pi)^2} \int d\mathbf{k}$, the renormalized Rabi frequency as a function of concentration (in lowest order of $n a_B^2$) reads
\begin{align}
\label{eq:G_linear}
G(n)= G(0) \left(1 - 2N \frac{\sum_{\mathbf{k}} |\phi_{\mathbf{k}}|^2 \phi_{\mathbf{k}}}{\sum_{\mathbf{k'}} \phi_{\mathbf{k'}}^*} + O[N^2] \right) = G(0) \left(1 - \frac{16\pi n a_B^2}{7} + O[n^2 a_B^4] \right),
\end{align}
where $G(0) = g \sqrt{2A/\pi a_B^2} = \frac{e p_{cv}}{m a_B^2} \sqrt{\frac{\hbar^2}{\pi \epsilon \epsilon_0 \omega_{\mathrm{cav}} L_{\mathrm{cav}}}}$.
Finally, we need to account for the fact that $G(n)$ is a derived term in the equations of motion, which includes $\propto n$ correction. Thus it originates from the effective nonlinear Hamiltonian $\hat{H}_{\mathrm{coupl}}^{\mathrm{(nonlin)}}$ which contains three excitonic operators and a photonic one, and on the contrary to the linear case provides extra factor of $2$ in the equations of motion. The modified Hamiltonian with exciton density-dependent Rabi frequency then reads (at fixed momentum)
\begin{align}
\label{eq:H_coupl_mod}
\hat{H}_{\mathrm{coupl}} = \frac{\Omega_X (n_X)}{2} (\hat{X}^\dagger \hat{c} + \hat{c}^\dagger \hat{X}),
\end{align}
where renormalized Rabi frequency is 
\begin{align}
\label{eq:Omega_R_linear}
\frac{\Omega_X (n_X)}{2}= \frac{\Omega_X^{(0)}}{2} \left(1 - \frac{8\pi n_X a_B^2}{7} + O[n_X^2 a_B^4] \right),
\end{align}
where we denoted the exciton concentration as $n_X$. The result above coincides with the estimates by Tassone and Yamamoto,\cite{Tassone1999} and Rochat et al.,\cite{Rochat2000} although derived in a different way, without involving Usui transformation.

We can proceed to calculate the terms being higher order in $(n a_B^2)$. This relies on the exact calculation of average as\cite{CombescotReview}
\begin{align}
\langle \text{\o} | \hat{X}_0^N \hat{X}_j^\dagger (\hat{X}_0^\dagger)^{N-1} | \text{\o}\rangle / \langle N|N\rangle = \delta_{0,j} \frac{F_{N-1}}{F_N} - (N-1) \lambda \left( \begin{array}{cc}
    0 & 0 \\
    0 & j \end{array} \right) \frac{F_{N-2}}{F_N} + O[n^3 a_B^6],
\end{align}
where $F_N$ is a coefficient which defines the deviation of statistics through $\langle N|N\rangle \equiv N! F_N$, with $F_N$ being $1$ for purely bosonic states. Exploiting coboson theory, the ratio reads
\begin{align}
\label{eq:FN_ratio}
\frac{F_{N-1}}{F_N} = 1 + N \sum_{\mathbf{k}} |\phi_{\mathbf{k}}|^4,
\end{align}
and $F_{N-2}/F_N \approx (F_{N-1}/F_N)^2$. The first term in Eq. \eqref{eq:FN_ratio} corresponds to the previously obtained case with $\sim n a_B^2$ scaling. Performing the same procedure as before, we extend the results to include $n^2 a_B^4$ contribution. After some algebra, we obtain
\begin{align}
\label{eq:G_full}
&G(n) = G(0) \left\{1 - 2N \frac{\sum_{\mathbf{k}} |\phi_{\mathbf{k}}|^2 \phi_{\mathbf{k}}}{\sum_{\mathbf{k'}} \phi_{\mathbf{k'}}^*} - 2N^2 \frac{(\sum_{\mathbf{k}} |\phi_{\mathbf{k}}|^4) (\sum_{\mathbf{k''}} |\phi_{\mathbf{k''}}|^2 \phi_{\mathbf{k''}})}{\sum_{\mathbf{k'}} \phi_{\mathbf{k'}}^*} + 2N^2 \frac{(\sum_{\mathbf{k}} |\phi_{\mathbf{k}}|^4 \phi_{\mathbf{k}})}{\sum_{\mathbf{k'}} \phi_{\mathbf{k'}}^*} + O[N^3] \right\} \\ \notag  &= G(0) \left\{1 - \frac{16\pi n a_B^2}{7} + \frac{1152\pi^2 n^2 a_B^4}{455} + O[n^3 a_B^6] \right\},
\end{align}
where the derivation was performed up to $\sim n^3 a_B^6$ terms. The corresponding modified Rabi frequency then reads
\begin{align}
\label{eq:Omega_R_full}
\frac{\Omega_X (n_X)}{2} = \frac{\Omega_X^{(0)}}{2} \left\{1 - \frac{8\pi n_X a_B^2}{7} + \frac{384\pi^2 n_X^2 a_B^4}{455} + O[n_X^3 a_B^6] \right\}.
\end{align}

Interestingly, we observe that the light-matter coupling strength is a monotonically decreasing function of density, and its expansion has a sign-changing pattern. It is thus tempting to suggest that it has the exponential dependence. Namely, we can express it as $G(n)=G(0)\exp\{- (16\pi/7) n a_B^2\} \approx G(0) \Big\{ 1 - (16\pi/7) n a_B^2 + (128\pi^2/49) n^2 a_B^4 + O[n^3 a_B^6] \Big\}$, and observe that the expansion of the exponent gives nearly the same quadratic term, with their ratio being $1.03175 \approx 1$. This gives us confidence in the conjectured dependence.\\

\subsection{Nonlinear exciton energy shift} In the previous section, we have considered the renormalization of the light-matter coupling term, which gains $n$-dependence. Now, let us consider similar effects which provide nonlinear energy term for composite excitons. To do so, we write the exciton Hamiltonian in terms of basic constituents, being electrons and holes. This is given by
\begin{align}
\label{eq:H_X_eh}
\hat{H}_{X} = \hat{H}_{e}^{(0)} + \hat{H}_{h}^{(0)} +  \hat{H}_{e-h} + \hat{H}_{e-e} + \hat{H}_{h-h},
\end{align}
where the first two terms correspond to energies of the electrons and holes, and read $\hat{H}_{e}^{(0)} = \sum_{\mathbf{k}_{\alpha}} \epsilon_{\mathbf{k}_{\alpha}}^{(e)} \hat{a}_{\mathbf{k}_{\alpha}}^\dagger \hat{a}_{\mathbf{k}_{\alpha}}$ and $\hat{H}_{h}^{(0)} = \sum_{\mathbf{k}_{\beta}} \epsilon_{\mathbf{k}_{\beta}}^{(h)}\hat{b}_{\mathbf{k}_{\beta}}^\dagger \hat{b}_{\mathbf{k}_{\beta}}$. The dispersions can be written in the quadratic form, being $\epsilon_{\mathbf{k}_{\alpha}}^{(e)} = E_c + k_{\alpha}^2/2m_e$ and $\epsilon_{\mathbf{k}_{\beta}}^{(h)} = E_v + k_{\beta}^2/2m_h$. $(E_c-E_v):= E_g$ corresponds to the bandgap energy, and $m_{e,h}$ are effective electron and hole masses, respectively, measured in units of free electron mass. The third term in Eq. \eqref{eq:H_X_eh} corresponds to electron-hole Coulomb interaction, which ultimately leads to the formation of the bound state. It reads $\hat{H}_{e-h} = - \sum_{\mathbf{p},\mathbf{p'},\mathbf{q}} V_q \hat{a}_{\mathbf{p}+\mathbf{q}}^\dagger \hat{b}_{\mathbf{p'}-\mathbf{q}}^\dagger \hat{b}_{\mathbf{p'}} \hat{a}_{\mathbf{p}}$, where $V_{q} = 2\pi e^2/(Aq S_q)$ is the standard Fourier transform for Coulomb interaction in 2D, $S_q$ denotes the screening function (so far unspecified), and we accounted explicitly for attraction between an electron and a hole. Finally, the terms $\hat{H}_{e-e}$ and $\hat{H}_{h-h}$ correspond to Coulomb interaction with only electrons and only holes. When accounting for the excitonic structure of e-h complexes, these lead to Kerr-type exciton-exciton interaction.

To account for the non-bosonicity and exchange-based Coulomb scattering between the cobosons on the equal footing, we can derive the total energy of the system, $E_X = \langle \hat{H}_X \rangle$, where expectation value is taken over $N$-exciton state. Following Ref. [\onlinecite{CombescotReview}], the nonlinear (quadratic and higher) contribution to the energy of $N$ excitons reads
\begin{align}
\label{eq:H_N_general}
\langle \hat{H}_X \rangle_N = &N \Bigg[ E^{(0)}_X + \frac{N}{2}  \frac{F_{N-2}}{F_N} \left\{ \xi \left( \begin{array}{cc}
    0 & 0 \\
    0 & 0 \end{array} \right) - \xi^{\mathrm{in}} \left( \begin{array}{cc}
    0 & 0 \\
    0 & 0 \end{array} \right) \right\} + \frac{N^2}{4}  \frac{F_{N-3}}{F_N} \Bigg\{ -2 \sum_n \lambda \left( \begin{array}{cc}
    0 & n \\
    0 & 0 \end{array} \right) \xi \left( \begin{array}{cc}
    n & 0 \\
    0 & 0 \end{array} \right) \\ \notag &+ \sum_{mn} \lambda \left( \begin{array}{cc}
    0 & 0 \\
    0 & n \\
    0 & m \end{array} \right) \xi \left( \begin{array}{cc}
    n & 0 \\
    m & 0 \end{array} \right) \Bigg\} \Bigg],
\end{align}
where $E^{(0)}_X = E_g - E_b$ is a density independent exciton energy, and nonbosonicity factor reads $F_{N-3}/F_N \approx (F_{N-1}/F_N)^3$. Here, 
\begin{align}
\label{eq:xi_direct}
\xi \left( \begin{array}{cc}
    n & j \\
    m & i \end{array} \right) = \int d\mathbf{r}_{\alpha_1} d\mathbf{r}_{\alpha_2} d\mathbf{r}_{\beta_1} d\mathbf{r}_{\beta_2} \phi_{m}^*(\mathbf{r}_{\alpha_1},\mathbf{r}_{\beta_1}) \phi_{n}^*(\mathbf{r}_{\alpha_2},\mathbf{r}_{\beta_2}) \phi_{i}(\mathbf{r}_{\alpha_1},\mathbf{r}_{\beta_1}) \phi_{j}(\mathbf{r}_{\alpha_2},\mathbf{r}_{\beta_2}) \\ \notag \times \left[ \mathcal{V}_{\alpha \alpha}(\mathbf{r}_{\alpha_1}, \mathbf{r}_{\alpha_2}) + \mathcal{V}_{\beta \beta}(\mathbf{r}_{\beta_1}, \mathbf{r}_{\beta_2}) + \mathcal{V}_{\alpha \beta}(\mathbf{r}_{\alpha_1}, \mathbf{r}_{\beta_2}) + \mathcal{V}_{\alpha \beta}(\mathbf{r}_{\alpha_2}, \mathbf{r}_{\beta_1})\right],
\end{align}
is a direct Coulomb scattering term between excitons [see Eq. \eqref{eq:lambda_gen} for comparison], and $\mathcal{V}_{f f'}(\mathbf{r}_{f_1}, \mathbf{r}_{f_2'})$ corresponds to the real space Coulomb potential between carriers $f, f'$. The $\xi^{\mathrm{in}}$ term denotes exchange Coulomb scatterings, where either electron or hole is swapped between two composite excitons,
\begin{align}
\label{eq:xi_in}
\xi^{\mathrm{in}} \left( \begin{array}{cc}
    n & j \\
    m & i \end{array} \right) = \int d\mathbf{r}_{\alpha_1} d\mathbf{r}_{\alpha_2} d\mathbf{r}_{\beta_1} d\mathbf{r}_{\beta_2} \phi_{m}^*(\mathbf{r}_{\alpha_1},\mathbf{r}_{\beta_2}) \phi_{n}^*(\mathbf{r}_{\alpha_2},\mathbf{r}_{\beta_1}) \phi_{i}(\mathbf{r}_{\alpha_1},\mathbf{r}_{\beta_1}) \phi_{j}(\mathbf{r}_{\alpha_2},\mathbf{r}_{\beta_2}) \\ \notag \times \left[ \mathcal{V}_{\alpha \alpha}(\mathbf{r}_{\alpha_1}, \mathbf{r}_{\alpha_2}) + \mathcal{V}_{\beta \beta}(\mathbf{r}_{\beta_1}, \mathbf{r}_{\beta_2}) + \mathcal{V}_{\alpha \beta}(\mathbf{r}_{\alpha_1}, \mathbf{r}_{\beta_2}) + \mathcal{V}_{\alpha \beta}(\mathbf{r}_{\alpha_2}, \mathbf{r}_{\beta_1})\right],
\end{align}
which combines Pauli scattering and Coulomb interaction. Finally, the exchange term between three composite excitons (where carriers are swapped but no Coulomb vertex is included) yields
\begin{align}
\label{eq:lambda_6}
\lambda \left( \begin{array}{cc}
    p & k \\
    n & j \\
    m & i \end{array} \right) = \int d\mathbf{r}_{\alpha_1} d\mathbf{r}_{\alpha_2} d\mathbf{r}_{\alpha_3} d\mathbf{r}_{\beta_1} d\mathbf{r}_{\beta_2} d\mathbf{r}_{\beta_3} \phi_{m}^*(\mathbf{r}_{\alpha_1},\mathbf{r}_{\beta_2}) \phi_{n}^*(\mathbf{r}_{\alpha_3},\mathbf{r}_{\beta_1}) \phi_{p}^*(\mathbf{r}_{\alpha_2},\mathbf{r}_{\beta_3}) \phi_{i}(\mathbf{r}_{\alpha_1},\mathbf{r}_{\beta_1}) \phi_{j}(\mathbf{r}_{\alpha_2},\mathbf{r}_{\beta_2}) \phi_{k}(\mathbf{r}_{\alpha_3},\mathbf{r}_{\beta_3}).
\end{align}
For composite excitons with $\mathcal{V}_{\alpha \beta}(r) = -\mathcal{V}_{\alpha \alpha}(r)$, which is true for the electron-hole potential, the direct term vanishes, $\xi \left( \begin{array}{cc}
    0 & 0 \\
    0 & 0 \end{array} \right) = 0$. This can be seen as well-known absence of direct contribution at zero exchanged momentum, valid both for III-V semiconductors and TMDs. Similarly, $\sum_n \lambda \left( \begin{array}{cc}
    0 & n \\
    0 & 0 \end{array} \right) \xi \left( \begin{array}{cc}
    n & 0 \\
    0 & 0 \end{array} \right) = 0$, and only exchange terms shall be accounted. They can be calculated as
\begin{align}
\label{eq:xi_in_calc}
\xi^{\mathrm{in}} \left( \begin{array}{cc}
    n & j \\
    m & i \end{array} \right) = 2 \sum_{\mathbf{k},\mathbf{k}'} V_{\mathbf{k}-\mathbf{k}'} \Big\{ |\phi_{\mathbf{k}}|^2 |\phi_{\mathbf{k}'}|^2 - |\phi_{\mathbf{k}}|^2 \phi_{\mathbf{k}}^* \phi_{\mathbf{k}'} \Big\}
\end{align}
and 
\begin{align}
\label{eq:six_exchange}
\sum_{mn} \lambda \left( \begin{array}{cc}
    0 & 0 \\
    0 & n \\
    0 & m \end{array} \right) \xi \left( \begin{array}{cc}
    n & 0 \\
    m & 0 \end{array} \right) = 2 \sum_{\mathbf{k},\mathbf{k}'} V_{\mathbf{k}-\mathbf{k}'} |\phi_{\mathbf{k}}|^4 \Big\{  |\phi_{\mathbf{k}'}|^2 - \phi_{\mathbf{k}}^* \phi_{\mathbf{k}'} \Big\}.
\end{align}
The sums \eqref{eq:xi_in_calc} and \eqref{eq:six_exchange} can be converted into intergrals, and evaluated numerically for the exciton wavefunction in the form \eqref{eq:phi_k}. As an important consequence of the monolayer structure of the TMD, the potential is chosen to be screened,\cite{Cudazzo2011,Berkelbach2013,Chernikov2014} and has the form
\begin{align}
\label{eq:Vq_TMD}
V_{\mathbf{q}} = \frac{2\pi e^2}{(4\pi \epsilon_0)\kappa q (1 + r_0 q/\kappa)},
\end{align}
where $e$ is an electron charge, $\epsilon_0$ is a vacuum permittivity (note that SI units are used), $r_0$ is a screening length, and $\kappa = (\epsilon_{s1}+\epsilon_{s2})/2$ is an average dielectric permittivity for substrates from two sides.\cite{Chernikov2014} This directly follows from the well-known Keldysh potential of the form
\begin{align}
\mathcal{V}_{ee}(\mathbf{r}) = \frac{e^2}{(4\pi \epsilon_0) r_0} \frac{\pi}{2} \bigg[ H_0\left(\frac{\kappa r}{r_0}\right) - Y_0\left(\frac{\kappa r}{r_0}\right) \bigg],
\end{align}
defined with the help of Struve and Bessel functions of the second kind, and shown for the case of two electrons.

Finally, collecting the terms up to $N^2$ order, the nonlinear energy of the excitonic mode (as appearing in the Hamiltonian) can be written as
\begin{align}
\label{eq:EX_n}
E_X(n_X) = E_0 + \frac{8}{\pi} \frac{e^2}{4\pi \epsilon_0 \kappa a_B} \mathcal{I}_4(r_0) n_X a_B^2 - \frac{128}{5} \frac{e^2}{4\pi \epsilon_0 \kappa a_B} \Big[5 \mathcal{I}_6(r_0) -2 \mathcal{I}_4(r_0) \Big] n_X^2 a_B^4,
\end{align}
where exchange integrals for dimensionless length ($x = r/a_B$) depend on the screening length $r_0$ and read
\begin{align}
\label{eq:exchange_int_1}
&\mathcal{I}_4(r_0) = \int_0^{\infty} \int_0^{2\pi}  \frac{dx dx' d\theta 2 \pi x x'}{\sqrt{x^2 + x'^2 - 
     2 x x' \cos\theta} (1 + \frac{r_0}{\kappa a_B} \sqrt{
       x^2 + x'^2 - 2 x x' \cos\theta})}
    \frac{(-1)}{(1 + x^2)^3} \left( \frac{1}{(1 + x'^2)^3} - 
     \frac{1}{(1 + x^2)^{3/2}} \frac{1}{(1 + x'^2)^{3/2}} \right) ,\\
&\mathcal{I}_6(r_0) = \int_0^{\infty} \int_0^{2\pi}  \frac{dx dx' d\theta 2 \pi x x'}{\sqrt{x^2 + x'^2 - 
     2 x x' \cos\theta} (1 + \frac{r_0}{\kappa a_B} \sqrt{
       x^2 + x'^2 - 2 x x' \cos\theta})}
    \frac{(-1)}{(1 + x^2)^6} \left( \frac{1}{(1 + x'^2)^3} - 
     \frac{1}{(1 + x^2)^{3/2}} \frac{1}{(1 + x'^2)^{3/2}} \right) ,
\end{align}
where in the case of TMDC materials the dimensionless parameter $r_0/(\kappa a_B)$ enters the integrals. We perform the calculations considering MoSe$_2$ on hBN, where $r_0 = 4$~nm \cite{Berkelbach2013}, $\kappa = (\epsilon_{s1} + \epsilon_{s2})/2 = 4$ for hBN substrates, and leaving $a_B$ as a tuning parameter.\\

\subsection{Exciton-polaritons at increasing density} Taking our previously derived results for the renormalization of coupling and exciton properties, let us translate it to the case of polaritons.\cite{Brichkin2011} In the cases where trion mode can be excluded out considerations (it is weakly coupled and/or largely detuning), we use the two coupled modes Hamiltonian, concentrating on the exciton-photon coupling. This is justified by the experimental data at large detuning, where the trion fraction at X-C anti-crossing is estimated to be small ($<2\%$). The normal modes of neutral exciton-polariton system read
\begin{align}
\label{eq:omega_pm}
E_{\pm}(n_X)= \frac{E_C + E_{X}(n_X)}{2} \pm \frac{1}{2} \sqrt{\Omega_X(n_X)^2 + [E_C - E_X(n_X)]^2},
\end{align}
where $E_-(n_X) \equiv E_{\mathrm{MPB}}(n_X)$ corresponds to the middle polariton mode we are interested in. 

We proceed with applying the presented theory to explain the nonlinear blueshift of the middle polariton branch as function of an exciton concentration, for the case where trion resonance is largely detuned. Using the coefficients derived above, and corresponding density dependences for the coupling $\Omega_{X}(n_X)$ and exciton energy $E_X(n)$ terms, we plot the nonlinear energy shift as function of concentration. The theoretical results are shown in Fig.~4(b,d) of main text by red solid curves. Taking the exciton Bohr radius as the only fitting parameter, we set it to be $a_B = 0.85$~nm, which allows to qualitatively the behavior of the system. Furthermore, we verify the obtained value performing the variational procedure to obtain exciton properties in MoSe$_2$ covered with hBN. This can be done using the standard procedure with screened Keldysh potential with $r_0 = 4$~nm, $\kappa = 4$, and the effective electron mass $m_e = 0.8 m_0$,\cite{Larentis2018} which was measured to be rather large in MoSe$_2$ and similar to the effective mass of the hole, taken $m_h = 0.84 m_0$. This leads to the reduced mass $\mu = 0.41 m_0$. Performing minimization, we get $a_B^{\mathrm{(calc)}} = 0.93$~nm and binding energy $E_b^{\mathrm{(calc)}} = 259$~meV. These values lie close to the fitted value and experimentally measured energy, respectively. (As a bonus, in the next section devoted to trions we explicitly show how the same result can be easily obtained in the momentum space.)

We note that the ability to reproduce measured energy shift of $E_{\mathrm{MPB}}$ is only possible once both Rabi frequency reduction and nonlinear interactions are considered, while otherwise failing to provide required scaling. Namely, the inclusion of Coulomb-based exchange can only explain the observed behavior (2~meV shift within at order of magnitude change for the density) for either largely increased exciton Bohr radius in TMDC, which is unlikely, or much higher concentration going into $\sim 10^{13}$~cm$^{-2}$ range. At the same time, if only Rabi renormalization is accounted, the saturation of nonlinear shift cannot be reproduced.


\section{Nonlinear trion-polaritons: theory}
\label{sec:T}

\subsection{Trion-polariton Rabi-splitting} We consider the system with an initial doping, and study the effects of light-matter coupling with a multiparticle bound state. In the MoSe$_2$ TMD this corresponds to a negatively charged exciton --- trion --- which is spectroscopically located 30 meV below the excitonic resonance. We aim to estimate of the Rabi-splitting change for the case of a trion. In the similar fashion, the deviation from ideal statistics changes the value of trion-photon coupling. However, we note that the strong light-matter coupling regime for trion is much less studied, and its treatment requires extra care.

We begin with the interaction between the cavity and the trion mode. The latter can be generally described by a composite creation operator which creates two electrons and a hole from the vacuum state, $\hat{a}_{\mathbf{k}_e, s_e}^\dagger  \hat{a}_{\mathbf{k}_{e'}, s_{e'}}^\dagger \hat{b}_{\mathbf{k}_h, s_h}^\dagger |\text{\o} \rangle$. Here, $\mathbf{k}_{e,e',h}$ are the momenta of the respective individual constituents (so-called carrier coordinates \cite{Combescot2003}), and $s_{e,e',h}$ are the spin indices. We note that the most favorable trion configuration in MoSe$_2$ monolayer is the singlet state with two electrons having anti-parallel spin.\cite{MonteCarlo} The operator corresponding to the creation of a singlet state can be written as\cite{Ramon2003}
\begin{align}
\label{eq:T_singlet}
\hat{T}_{\mathbf{K},\uparrow}^\dagger = \sum_{\mathbf{k}_1, \mathbf{k}_2} \phi^{T}_{\beta_e \mathbf{K}- \mathbf{k}_1, \beta_e \mathbf{K}- \mathbf{k}_2} \frac{(\hat{a}_{\mathbf{k}_1 ,\uparrow}^\dagger \hat{a}_{\mathbf{k}_2, \downarrow}^\dagger - \hat{a}_{\mathbf{k}_2, \downarrow}^\dagger \hat{a}_{\mathbf{k}_1 ,\uparrow}^\dagger)}{\sqrt{2}}   \hat{b}_{\mathbf{K} -\mathbf{k}_1 - \mathbf{k}_2, \uparrow}^\dagger,
\end{align}
with wavefunction being separated into a trivial center-of-mass part with momentum $\mathbf{K}$, $\beta_e = 1 - \beta_X = m_e/(2m_e + m_h)$, and the relative motion part described by trial wavefunction for the relative motion $\phi^T_{\mathbf{k}_1,\mathbf{k}_2}$ (we consider zero CM momentum case). The wavefunction is written for the relative electron-hole coordinates $\mathbf{r}_1$ and $\mathbf{r}_2$, being radius-vectors between the first electron and the hole, and the second electron and the hole, respectively. In the real space it corresponds to the two exponentially decaying functions
\begin{align}
\label{eq:Phi_r12}
\phi^T(\mathbf{r}_1,\mathbf{r}_2) = \frac{1}{\sqrt{2}} \frac{1}{\sqrt{1 + \chi^2}} \bigg\{ \sqrt{\frac{2}{\pi \lambda_1^2}} e^{-r_1/\lambda_1} \sqrt{\frac{2}{\pi \lambda_2^2}} e^{-r_2/\lambda_2} + \sqrt{\frac{2}{\pi \lambda_2^2}} e^{-r_1/\lambda_2} \sqrt{\frac{2}{\pi \lambda_1^2}} e^{-r_2/\lambda_1} \bigg\},
\end{align}
where $\lambda_1$ and $\lambda_2$ are the variational parameters corresponding to the distances between electrons and a hole. Note, that $\phi^T(\mathbf{r}_1,\mathbf{r}_2)$ is symmetrized and is normalized to unity, with $\chi = 4 \lambda_1 \lambda_2/(\lambda_1 + \lambda_2)^2$. The momentum space version then reads
\begin{align}
\label{eq:Phi_k12}
&\phi^T_{\mathbf{k}_1,\mathbf{k}_2} = \frac{1}{\sqrt{2}} \frac{1}{\sqrt{1 + \chi^2}} \bigg\{ \sqrt{\frac{8\pi \lambda_1^2}{A}} (1 + \lambda_1^2 k_1^2)^{-3/2} \sqrt{\frac{8\pi \lambda_2^2}{A}} (1 + \lambda_2^2 k_2^2)^{-3/2} + \sqrt{\frac{8\pi \lambda_2^2}{A}} (1 + \lambda_2^2 k_1^2)^{-3/2} \sqrt{\frac{8\pi \lambda_1^2}{A}} (1 + \lambda_1^2 k_2^2)^{-3/2} \bigg\} \\ \notag &\equiv \mathcal{N} \Big\{ \phi_{\mathbf{k}_1}^{(1)} \phi_{\mathbf{k}_2}^{(2)} + \phi_{\mathbf{k}_1}^{(2)} \phi_{\mathbf{k}_2}^{(1)} \Big\},
\end{align}
where we defined $\mathcal{N} = [2 (1+\chi^2)]^{-1/2}$ and $\phi_{\mathbf{k}}^{(j)} = \sqrt{8\pi \lambda_j^2 /A} (1+ \lambda_j^2 k^2)^{-3/2}$. Finally, reordering the electron operators in Eq. \eqref{eq:T_singlet}, and considering CM momentum much smaller then typical relative momenta, we can write trion creation operator as
\begin{align}
\label{eq:T_dagger}
\hat{T}_{\mathbf{K},\uparrow}^\dagger = \sum_{\mathbf{k}_1, \mathbf{k}_2} \mathcal{N} \Big\{ \phi_{\mathbf{k}_1}^{(1)} \phi_{\mathbf{k}_2}^{(2)} + \phi_{\mathbf{k}_1}^{(2)} \phi_{\mathbf{k}_2}^{(1)} \Big\} \hat{a}_{\mathbf{k}_1 ,\uparrow}^\dagger \hat{a}_{\mathbf{k}_2, \downarrow}^\dagger \hat{b}_{\mathbf{K} -\mathbf{k}_1 - \mathbf{k}_2, \uparrow}^\dagger.
\end{align}

The choice of the wavefunction \eqref{eq:Phi_k12} is of course far from optimal, as to describe quantitatively the shape of trion solution, more complicated ansatzes with hundreds of orbitals shall be used.\cite{Kidd2016} However, in order to get any sensible result for Rabi frequency renormalization, this is the form we shall adopt.

Once there is a non-zero number of free electrons, the absorption of a circularly polarized photon can then allow a creation of a trion. The Hamiltonian of the system can be written as the sum $\hat{\mathcal{H}}^T = \hat{\mathcal{H}}^T_0 + \hat{\mathcal{H}}^T_{\mathrm{coupl}}$ of non-interaction cavity/electron/trion Hamiltonian $\hat{\mathcal{H}}^T_0 $
\begin{align}
\label{eq:H_trion_0}
\hat{\mathcal{H}}^T_0 = \sum_{\mathbf{q}} \omega_{\mathrm{cav},\mathbf{q}} \hat{c}^\dagger_{\mathbf{q}} \hat{c}_{\mathbf{q}} + \sum_{\mathbf{K}} \omega^T_{\mathbf{K}} \hat{T}_{\mathbf{K}}^\dagger \hat{T}_{\mathbf{K}} + \sum_{\mathbf{k}} \varepsilon_{\mathbf{k}} \hat{a}_{\mathbf{k}}^\dagger \hat{a}_{\mathbf{k}},
\end{align}
and the coupling Hamiltonian for light and matter,
\begin{align}
\label{eq:H_trion_int}
\hat{\mathcal{H}}^T_{\mathrm{coupl}} =  \sum_{\mathbf{k},\mathbf{q},\mathbf{k}_1,\mathbf{k}_2} g \phi^{T}_{\mathbf{k}_1,\mathbf{k}_2} \hat{T}_{\mathbf{k}+\mathbf{q},\uparrow}^\dagger \hat{a}_{\mathbf{k},\downarrow} \hat{c}_{\mathbf{q},\Uparrow} + h.c.,
\end{align}
with $g$ being conduction-to-valence band transition matrix element, previously defined in the exciton case. $\varepsilon_{\mathbf{k}}$ is an electron dispersion, $\omega^T_{\mathbf{K}}$ is a trion dispersion, and in Eq. \eqref{eq:H_trion_0} the summation over spin is assumed. In Eq. \eqref{eq:H_trion_int}, similarly to the exciton case, the wavefunction of the relative motion appears due to the fact that out of free electron-hole complex the bound trion state appears. The process in Eq. \eqref{eq:H_trion_int} in simple terms can be seen as a creation of electron-hole pair attached to the electron in a Fermi sea, while the electron state is (slightly) changed. This can be conveniently described by the quasi-bosonic excitation, defined by the operator $\hat{B}_j$, which reads 
\begin{align}
\label{eq:B_K}
\hat{B}_{\mathbf{K}}^\dagger |\mathrm{FS}\rangle = \frac{1}{\sqrt{N_e}} \sum_{\mathbf{k}} \hat{T}_{\mathbf{K} + \mathbf{k}}^\dagger \hat{a}_{\mathbf{k}} |\mathrm{FS}\rangle.
\end{align}
It creates an excitation out of Fermi sea state $|\mathrm{FS}\rangle$, and $N_e$ is a number of free electrons available for trion creation, which can correspond to the selected spin configuration, meaning that the total number of electrons in the system reads as $N_e^{\mathrm{tot}} = 2 N_e$. The combinatorial prefactor $1/\sqrt{N_e}$ comes from the number of different ways the excitation can be created,\cite{Rapaport2001} and we note that \eqref{eq:B_K} holds for low temperatures where electron gas is degenerate. As the excitation operator $\hat{B}_{\mathbf{K}}^\dagger$ represents a composite boson, similarly to excitons described in the previous section B, it is prone to the phase space filling effects. At the same time, it is not a bound state, and thus exhibits different statistics deviation behaviour. We will describe this point in details later, when trion-based saturation effects are considered.

We note that previously the light-matter coupling in MoSe$_2$ TMD material was also considered for the case of exciton-polarons.\cite{Sindler2017} This corresponds to similar creation operator $\hat{B}_{j}^\dagger$, but different ansatz for the wave function, which accounts for dressing of photo-created exciton with electrons in the Fermi sea. We stress that both trion-dominated or polaron-dominated regimes can be possible, as considered in Ref.~[\onlinecite{Shiau2017}]. Recent predictions for the GaAs samples estimate the cross-over into exciton-polaron regime to happen for Fermi wave vector to be comparable with inverse Bohr radius for the system, $k_{cr} \sim 0.8 a_B^{-1}$.\cite{Chang2018} As we show later, the experiment is conducted in the $k_F \ll a_B^{-1}$ limit of small concentration, which corresponds to the trion-dominated regime.

Next, we proceed with the estimation of trion Rabi-splitting, or conversely the free electron density. It is given by the bare coupling constant $g$ multiplied by the square root of electron density and the wave function part responsible for absorption renormalization due to the confinement. First, we account for the fact that photon momentum has typically small values $q$, being much less than other relevant wavevectors. This also translates into nearly zero center-of-mass momentum of the generated exciton, where an electron and a hole are located close to each other. Here, we follow the approach introduced in Refs. [\onlinecite{Shiau2012, Combescot2003b}], where the so-called electron-exciton coordinates are used. These correspond to the electron-hole relative coordinate (seen as an exciton) and the relative coordinate of the CM for ``exciton'' to the second electron. They are generally described by length parameters $\lambda$ and $\lambda'$. Due to complex symmetrization requirements, they do not allow to choose wavefunction in the simple form, thus preventing the analytical calculation. However, in the limit of large $\lambda_2 \gg \lambda_1$ the e-h and e-X coordinates become nearly equivalent, and we can set $\lambda \approx \lambda_1$ and $\lambda' \approx \lambda_2$.

Taking the trion wavefunction to be Fourier transformed with respect to an exciton internal motion,
\begin{align}
\label{eq:Phi_rk}
\phi^T_{\mathbf{k}_1,\mathbf{k}_2} = \int d\mathbf{r}_1 \frac{e^{i \mathbf{k}_1\cdot \mathbf{r}_1}}{\sqrt{A}} \mathcal{N} \bigg\{ \sqrt{\frac{2}{\pi \lambda_1^2}} \exp(-r_1/\lambda_1) \sqrt{\frac{8\pi \lambda_2^2}{A}} (1 + \lambda_2^2 k_2^2)^{-3/2} + \sqrt{\frac{2}{\pi \lambda_2^2}} \exp(-r_1/\lambda_2) \sqrt{\frac{8\pi \lambda_1^2}{A}} (1 + \lambda_1^2 k_2^2)^{-3/2} \bigg\},
\end{align}
the coupling then can be estimated setting $r_1=0$ (i.e. for the closely located photocreated e-h pair). Simultaneously, we shall account that Fermi wavevector $k_F \ll \lambda_1^{-1},\lambda_2^{-1}$, and thus the wavefunction in Eq. \eqref{eq:H_trion_int} can be considered as a constant at $k_2\lambda_{2,1}=0$, going in front of the sum. Altogether, the coupling Hamiltonian can be rewritten as
\begin{align}
\label{eq:H_trion_Rabi}
\hat{\mathcal{H}}^T_{\mathrm{coupl}} =  \frac{\Omega_T}{2} \sum_{\mathbf{q}} (\hat{B}^\dagger_{\mathbf{q}} \hat{c}_{\mathbf{q}} + h.c.),
\end{align}
where the trion Rabi frequency reads
\begin{align}
\label{eq:trion_Rabi}
\frac{\Omega_T}{2} = g \sqrt{N_e} 4\mathcal{N} \left( \frac{\lambda_2}{\lambda_1} + \frac{\lambda_1}{\lambda_2} \right) = \frac{\Omega_X}{2} \sqrt{8 \pi a_B^2} \mathcal{N} \left( \frac{\lambda_2}{\lambda_1} + \frac{\lambda_1}{\lambda_2} \right) \sqrt{n_e},
\end{align}
where we used the Rabi frequency definition for the neutral exciton case, $\Omega_X/2 = g \sqrt{2A/\pi a_B^2}$, and $n_e = N_e/A$ is a concentration of free electrons which can form trions. Eq.~\eqref{eq:trion_Rabi} then allows to estimate $n_e$ using
\begin{align}
\label{eq:ne}
n_e = \left(\frac{\Omega_T}{\Omega_X}\right)^2 \frac{1}{4\pi a_B^2} \frac{[1 + 16\lambda_1^2 \lambda_2^2 / (\lambda_1 + \lambda_2)^4]}{(\lambda_2/\lambda_1 + \lambda_1/\lambda_2)^2},
\end{align}
once the variational parameters are known.

\subsection{Trion binding energy and variation} Next, we proceed to define $\lambda_1$ and $\lambda_2$ for trions in TMDC. We follow the approach outlined in Ref. [\onlinecite{Ramon2003}], using the wavefunction \eqref{eq:T_dagger}. The expectation value for the trion Hamiltonian (includes kinetic terms for relative motion and Coulomb interaction) then can be written as
\begin{align}
\label{eq:E_T_gen}
E^{(T)} = \frac{\mathcal{I}_1 + 2\chi \mathcal{I}_2 + \mathcal{J}_1 + \mathcal{J}_2}{1+\chi^2},
\end{align}
where we define auxiliary quantities:
\begin{align}
\label{eq:I1}
&\mathcal{I}_1 = \left( \frac{1}{\lambda_1^2} + \frac{1}{\lambda_2^2}\right) (1+\gamma) -\frac{4}{\lambda_1}\int_0^{\infty} dx \frac{1}{(1+ \frac{2 r_0}{\kappa \lambda_1} x)} \frac{1}{[1+ x^2]^{3/2}} -\frac{4}{\lambda_2}\int_0^{\infty} dx \frac{1}{(1+ \frac{2 r_0}{\kappa \lambda_2} x)} \frac{1}{[1+ x^2]^{3/2}},\\
\label{eq:I2}
&\mathcal{I}_2 = \frac{4 (1+\gamma)}{(\lambda_1 + \lambda_2)^2} -\frac{(\lambda_1 + \lambda_2) \chi^2}{2 \tilde{\lambda}^2} \int_0^{\infty} dx \frac{1}{(1+ \frac{ r_0}{\kappa \tilde{\lambda}} x)} \frac{1}{[1+ x^2]^{3/2}} ,\\
&\mathcal{J}_1 = \frac{4}{\lambda_1} \int_0^{\infty} dx \frac{1}{(1+ \frac{2 r_0}{\kappa \lambda_1} x)} \frac{1}{[1+ x^2]^{3/2}} \frac{1}{[1+ (\lambda_2/\lambda_1)^2 x^2]^{3/2}},\\
&\mathcal{J}_2 = \frac{2 \chi^2}{\tilde{\lambda}} \int_0^{\infty} dx \frac{1}{(1+ \frac{r_0}{\kappa \tilde{\lambda}} x)} \frac{1}{[1+ x^2]^{3}} ,
\end{align}
where $\gamma=m_e/m_h$, $\tilde{\lambda} = \lambda_1 \lambda_2/(\lambda_1 + \lambda_2)$, and we remind that $r_0$ is screening parameter, $\kappa$ is average dielectric permittivity of the substrate. Here, all length parameters are measured in the units of 
\begin{align}
\label{eq:a0}
a_0 = \frac{\hbar^2 \epsilon_0 4 \pi \kappa}{e^2 m_e},
\end{align}
and energies are measured in units of
\begin{align}
\label{eq:E0}
E_0 = \frac{\hbar^2}{2 m_e a_0^2}.
\end{align}
To obtain the binding energy for the trion complex, we minimize $E^{(T)}[\lambda_1,\lambda_2]$ with respect to variational parameters, and subtract the exciton binding energy contribution $E_b^{X}$. The latter is obtained from minimization of 
\begin{align}
E^{(X)}[\lambda_0] = \frac{(1+\gamma)}{\lambda_0^2} - \frac{4}{\lambda_0} \int_0^{\infty} dx \frac{1}{(1+ \frac{2 r_0}{\kappa \lambda_0} x)} \frac{1}{[1+ x^2]^{3/2}},
\end{align}
which for previously defined parameters of $m_e = 0.8 m_0$, $m_h = 0.84 m_0$, $r_0 = 4$~nm, $\kappa = 4$, gives $\lambda_0 = 0.93$~nm and binding energy of $E_b^{X} = -\min \{E^{(X)}[\lambda_0] \} = 259$~meV. The variational procedure for the trion then gives the binding energy of $E_b^{T} = -\min \{E^{(T)}[\lambda_1, \lambda_2] \} - E_b^{X} = 26$~meV for $\lambda_1 = 0.87$~nm and $\lambda_2 = 2.54$~nm. These are the parameters which will be used in the following. Although we remind that considered variation with two parameters is oversimplistic, it provides energy estimate to be very close experimentally measured trion binding energy of $30$~meV. 

Finally, substituting obtained radii $\lambda_{1,2}$, $a_B = 0.93$~nm, and experimentally measured Rabi-splittings $\Omega_T = 5.8$~meV, $\Omega_X=17.2$~meV, using Eq. \eqref{eq:ne} we estimate the electron concentration available for trion creation to be $n_e = 4.05 \times 10^{10}$~cm$^{-2}$, with the full concentration corresponding to $n_e^{\mathrm{tot}} = 8.1 \times 10^{10}$~cm$^{-2}$. Note that so far only variation and parameters obtained from the exciton-polariton case were used, with no fitting involved.

\subsection{Trion Rabi-splitting quench} We continue with the calculation of the modified trion Rabi frequency due to the deviation of statistics. For this, similarly to excitonic case [Eq. \eqref{eq:dXdt}], we derive the equations of motion for the excitation mode $\hat{B}_j$ using Hamiltonian \eqref{eq:H_trion_Rabi}. The nontrivial dynamics part comes from the light-matter coupling term
\begin{align}
\label{eq:dBdt}
i\frac{d \hat{B}_{\mathbf{q'}}}{d t} \Big|_{\mathrm{coupl}} = [\hat{B}_{\mathbf{q}'}, \hat{\mathcal{H}}^T_{\mathrm{coupl}}] = \frac{\Omega_T}{2} \sum_{\mathbf{q}} \hat{c}_{\mathbf{q}} [\hat{B}_{\mathbf{q}'}, \hat{B}_{\mathbf{q}}^\dagger],
\end{align}
and relies on the calculation of commutator $[\hat{B}_{\mathbf{q}'}, \hat{B}_{\mathbf{q}}^\dagger]$. To do so, it is instructive to rewrite the excitation operator in terms of trion and electron, yielding 
\begin{align}
\label{eq:comm_B}
\Big[\hat{B}_{\mathbf{q}'}, \hat{B}_{\mathbf{q}}^\dagger \Big] = \frac{1}{N_e} \bigg[\sum_{\mathbf{k}'} \hat{a}^\dagger_{\mathbf{k}'} \hat{T}_{\mathbf{q}'+\mathbf{k}'} , \sum_{\mathbf{k}} \hat{T}^\dagger_{\mathbf{q}+\mathbf{k}} \hat{a}_{\mathbf{k}} \bigg] = \frac{1}{N_e} \sum_{\mathbf{k},\mathbf{k}'} \bigg(  \hat{a}^\dagger_{\mathbf{k}'} \Big\{ \hat{T}_{\mathbf{q}'+\mathbf{k}'} , \hat{T}^\dagger_{\mathbf{q}+\mathbf{k}} \Big\} \hat{a}_{\mathbf{k}} - \hat{T}^\dagger_{\mathbf{q}+\mathbf{k}} \hat{T}_{\mathbf{q}'+\mathbf{k}'} \delta_{\mathbf{k},\mathbf{k}'} \bigg).
\end{align}
In the case of low pumping intensity, the trion anti-commutation relations resemble that of ideal fermions, $\{ \hat{T}_i , \hat{T}^\dagger_j \} = \delta_{i,j}$, and the number of trions goes to zero. Then, the commutator for $\mathbf{q} = \mathbf{q}'$ reduces to integral over distribution function $f_{\mathbf{k}}$ and gives unity,
\begin{align}
\label{eq:comm_B=1}
\Big[\hat{B}_{\mathbf{q}}, \hat{B}_{\mathbf{q}}^\dagger \Big] = \frac{1}{N_e} \sum_{\mathbf{k}} f_{\mathbf{k}} = 1,
\end{align}
as it should be for an ideal bosonic mode. However, for the increase of pumping we observe two contributions which change the commutation relation. First contribution comes from deviation of fermionicity for composite trion operator, such that $\{ \hat{T}_{\mathbf{q}'+\mathbf{k}'} , \hat{T}^\dagger_{\mathbf{q}+\mathbf{k}} \}$ is not a simple delta function anymore. For equal momenta this starts at the value of unity, and decreases with powers of $n_T \lambda^2$. 

The second contribution comes from the composite nature of the quasi-bosonic operator $\hat{B}_{\mathbf{q}}$, which ultimately depends on the ratio between number of trions and available free electrons for their creation. To demonstrate this point, let us rewrite Eq.~\eqref{eq:comm_B} as
\begin{align}
\label{eq:comm_B_dev}
\Big[\hat{B}_{\mathbf{q}'}, \hat{B}_{\mathbf{q}}^\dagger \Big] = \delta_{\mathbf{q}, \mathbf{q'}} - \hat{D}_{\mathbf{q}, \mathbf{q'}},
\end{align}
where the deviation operator is formally introduced as 
\begin{align}
\label{eq:D_trions}
\hat{D}_{\mathbf{q}, \mathbf{q'}} = \delta_{\mathbf{q}, \mathbf{q'}} - \frac{1}{N_e} \sum_{k < k_{\mathrm{F}}} \bigg(  \hat{a}^\dagger_{\mathbf{k} + \mathbf{q} - \mathbf{q}'} \hat{a}_{\mathbf{k}} - \hat{T}^\dagger_{\mathbf{q}+\mathbf{k}} \hat{T}_{\mathbf{q}'+\mathbf{k}} \bigg),
\end{align}
where we have considered the limit of $n_T \lambda^2 \ll 1$, such that trions can be approximately treated as ideal fermions. We observe that while conceptually the deviation operator resembles the one used for excitons in the previous section, the fact that trion-electron excitation is not bound leads to different closure relations and statistics. In this case, it is reminiscent to intersubband excitations\cite{DeLiberato2009} with trion being an excitation over the Fermi sea. We proceed by deriving the commutation relations for the devation operator and excitation operator, which reads
\begin{align}
\label{eq:D_trions_comm}
\Big[\hat{D}_{\mathbf{q}, \mathbf{q'}}, \hat{B}_{\mathbf{q}''}^\dagger \Big] = \frac{2}{N_e}  \hat{B}_{\mathbf{q}'' + \mathbf{q} - \mathbf{q}'}^\dagger,
\end{align}
and it can be recursively generalized to the case of $N_T$ particles as
\begin{align}
\label{eq:D_trions_comm_NT}
\Big[\hat{D}_{\mathbf{q}, \mathbf{q'}}, (\hat{B}_{\mathbf{q}''}^\dagger)^{N_T} \Big] = \frac{2 N_T}{N_e}  (\hat{B}_{\mathbf{q}''}^\dagger)^{N_T - 1} \hat{B}_{\mathbf{q}'' + \mathbf{q} - \mathbf{q}'}^\dagger.
\end{align}
The derived commutation relations, which are dependent on $N_T/N_e$ ratio, will be later shown to ultimately lead to the quench of trion Rabi frequency, where the commutator $\langle \Big[\hat{B}_{\mathbf{q}'}, \hat{B}_{\mathbf{q}}^\dagger \Big] \rangle$ averaged over highly-excited many-body state vanishes.

In the following, we proceed with considering the two aforementioned contributions one-by-one. \\

\subsection{Deviation from fermionicity for the composite trion anti-commutator} To account for the deviation of statistics, we use the generalized many-body formalism for composite n-particles \cite{Combescot2010}. This allows to calculate anti-commutator for the composite fermion (trion in our case), which consists of three particles of different flavor (opposite spin electrons an a hole). In the general form, it reads
\begin{align}
\label{eq:anticomm_T}
\Big\{ \hat{T}_m, \hat{T}_i^\dagger \Big\} = \delta_{m,i} - \hat{\Xi}_{mi},
\end{align}
where deviation from fermionicity operator $\hat{\Xi}_{mi}$ is defined as
\begin{align}
\label{eq:Theta_mi}
\Big[ \hat{\Xi}_{mi}, \hat{T}_j^\dagger \Big] = \sum_n \hat{T}_n^\dagger \sum_\rho \left( \lambda_\rho \left( \begin{array}{cc}
    n & j \\
    m & i \end{array} \right) - \lambda_\rho \left( \begin{array}{cc}
    n & i \\
    m & j \end{array} \right) \right) + \hat{\Xi}_{mij}^\dagger,
\end{align}
and the operator $\hat{\Xi}_{mij}^\dagger$ is defined through the anticommutator
\begin{align}
\label{eq:Theta_mij}
\Big\{ \hat{\Xi}_{mij}^\dagger, \hat{T}_k^\dagger \Big\} = \sum_{p,n} \hat{T}_p^\dagger \hat{T}_n^\dagger \Lambda_{p,k,n,j,m,i}.
\end{align}
Here single exchange integrals $\lambda_\rho(j,i,n,m)$ are as in Eq. \eqref{eq:lambda_gen} (though with three particle wavefunction), and $\rho$ denotes the carrier to be exchanged. In total, it provides six contributions, which in the case of zero exchanged momentum we expect to be the same. The last term in \eqref{eq:Theta_mi} corresponds to three-particle exchanges $\Lambda_{p,k,n,j,m,i}$ with all permutations. As it typically corresponds to $\sim \lambda^4$ scaling, which shall be accompanied by the quadratic density contribution, we refrain from considering it, and concentrate on lower order terms only. However, we note that the missing term might still effect the quench of a trion Rabi, as at increasing concentration the terms in all orders become important.

The trion exchange is calculated using the trial wavefunction 
\begin{align}
\label{eq:Phi_non-sym}
\Phi^T_{\mathbf{p},\mathbf{k}} = \sqrt{\frac{8\pi \lambda^2}{A}} (1 + \lambda^2 p^2)^{-3/2} \sqrt{\frac{8\pi \lambda'^2}{A}} (1 + \lambda'^2 k^2)^{-3/2}
\end{align}
in the electron-exciton basis without symmetrization, which is accounted at latter stage, and we take hole exchange as an example. In analogy to the case of trion interaction\cite{Combescot2011} it reads
\begin{align}
\label{eq:lambda_T}
\lambda_h \left( \begin{array}{cc}
    0 & 0 \\
    0 & 0 \end{array} \right) = \sum_{\mathbf{k},\mathbf{p},\mathbf{p}'}  |\Phi^T_{\mathbf{p},\mathbf{k}}|^2 |\Phi^T_{\mathbf{p}',\mathbf{k}+\alpha_h (\mathbf{p} - \mathbf{p}')}|^2 =: \lambda_T.
\end{align}
To evaluate \eqref{eq:lambda_T} we use new coordinates $\mathbf{p} - \mathbf{p}' = \mathbf{\delta p}, (\mathbf{p} + \mathbf{p}')/2 = \mathbf{P}$, make momenta dimensionless multiplying it by $\lambda$, and define $\xi = \lambda'/\lambda$. Finally, following the same procedure as for composite excitons, the anti-commutator $\langle N_T| \{ \hat{T}_{\mathbf{q}'}, \hat{T}_{\mathbf{q}}^\dagger \} |N_T\rangle$ is averaged over a state of $N_T$ composite particles (now \emph{fermions}), such that $N_T$ contributions is obtained. This leads to the estimate for the deviation
\begin{align}
\label{eq:B_comm_nF_calc}
\langle \Big[\hat{B}_{\mathbf{q}}, \hat{B}_{\mathbf{q}}^\dagger \Big]_{nF} \rangle \equiv  \frac{1}{N_e} \langle \sum_{\mathbf{k},\mathbf{k}'}  \hat{a}^\dagger_{\mathbf{k}'} \hat{\Xi}_{\mathbf{q}'+\mathbf{k}',\mathbf{q}+\mathbf{k}}  \hat{a}_{\mathbf{k}} \rangle = - \bigg\{ 6 \times \frac{128}{\pi} \left( \frac{\lambda'^4}{\lambda^2} \right) \frac{N_T}{A} \mathcal{I}_T + (\lambda \leftrightarrow \lambda') \bigg\} =: -\Delta_{nF},
\end{align}
where the dimensionless exchange integral reads
\begin{align}
\label{eq:I_exch_T}
\mathcal{I}_T = &\int_0^{\infty} \int_0^{2\pi} dx dx' dy d\theta_1 d\theta_2 x x' y \frac{1}{(1 + x^2 + x'^2/4 + 
     x x' \cos\theta_1 )^3 }
    \frac{1}{(1 + \xi^2 y^2)^3} \frac{1}{(1 + x^2 + x'^2/4 - 
     x x' \cos\theta_1 )^3 } \\ &\times \frac{1}{(1 + \xi^2 [y^2 + x'^2/4 - 
     y x' \cos\theta_2] )^3 } ,
\end{align}
and we account for symmetrization between $\lambda$ and $\lambda'$ parameter as a separate term. As in the case of excitons, we observe that first order correction in $n_T \tilde{\lambda}^2$ reduces the coupling as a function of composite particle density ($\tilde{\lambda}$ is an effective parameter of length dimensionality).\\

\subsection{Effects of medium saturation} Next, we find that the increased number of trions as a consequence of increasing pump intensity provides another contribution for the trion Rabi frequency reduction. To derive the trion-density dependence for $\Omega_T$, we employ the same strategy as previously used for composite excitons in Sec. B.1. For this, we consider the ground state of the system as a Fermi sea of free electrons $| \mathrm{FS} \rangle$ available for the trion creation. The relevant excited states then correspond to multi-trion states $|N_T \rangle \equiv (\hat{B}_{\mathbf{q}''}^\dagger)^{N_T} |\mathrm{FS}\rangle$. The nonlinear contribution to the trion Rabi frequency associated to the composite nature of $\hat{B}_{\mathbf{q}''}$ comes from the average
\begin{align}
\label{eq:BB_comm_Rabi}
\langle N_T | \Big[\hat{B}_{\mathbf{q}'}, \hat{B}_{\mathbf{q}}^\dagger \Big] |N_T\rangle = \langle N_T | N_T\rangle - \langle \mathrm{FS} | \hat{B}_{\mathbf{q}''}^{N_T} \Big[\hat{D}_{\mathbf{q},\mathbf{q}'}, \hat{B}_{\mathbf{q}''}^{\dagger N_T} \Big] |\mathrm{FS}\rangle \approx 1 - \frac{2N_T}{N_e} + O[(N_T/N_e)^2],
\end{align}
where we used Eq.~\eqref{eq:D_trions_comm_NT} and the fact that $q,q',q''$ are small. The analysis of Eq.~\eqref{eq:BB_comm_Rabi} shows that modified commutation relations can ultimately leads to the quench of the strong coupling once the number of trions becomes comparable to half the number of free electrons $N_T = N_e/2$. However, this only corresponds to the lowest order corrections, and higher terms shall be accounted for increasing $N_T/N_e$ ratio to get the full treatment. In this case, smooth reduction of $\Omega_T$ is expected up to $N_T = N_e$.


Finally, collecting all contributions together (including the one described in Sec. C.4), we can write the effective commutator at growing trion density $n_T = N_T/A$ as a function
\begin{align}
\label{eq:B_comm_together}
\langle \Big[\hat{B}_{\mathbf{q}}, \hat{B}_{\mathbf{q}}^\dagger \Big] \rangle = 1 - \frac{2 n_T}{n_e} -\Delta_{nF} + \Delta_{nF}^2/2 + O[n_T \tilde{\lambda}^2] := f_T (n_T, n_e, \lambda, \lambda'),
\end{align}
where similarly to exciton case we conjectured the appearance of the quadratic term $\Delta_{nF}^2/2$, which appears in the expansion of the exponent. The function $f_T(N_T, N_e, \lambda, \lambda')$ is decreasing from 1 to 0, and we consider it zero after the quench. The important parameter then is the half of available electron density $n_e/2$, which defines the excitation density at which quench is observed.
The resulting density dependent trion Rabi frequency, defined as in Eq.~\eqref{eq:dBdt}, then reads
\begin{align}
\frac{\Omega_T (n_T)}{2} = \frac{\Omega_T (0)}{2} \left( 1 - \frac{2 n_T}{n_e} -\Delta_{nF} + \Delta_{nF}^2/2 \right),
\end{align}
and is used later to calculate the density dependence for the polariton modes.

\textit{Intuitive explanation.} To explain the leading trend of linearly decreased coupling for the number of trion equal to half the number of free electrons, we note that the coupling of the trion mode to the cavity bares analogy to the atom-photon coupling\cite{Rapaport2001,Rapaport2000} (contrary to the neutral exciton case). This is readily seen in the $\propto \sqrt{n_e}$ dependence for the coupling constant, similarly to the common square root enhancement for the $N$ two-level emitters.\cite{PolzikRev} Given this correspondence, we show how the coupling between trion and photon changes for high excitation power. For the trion case, the state $|g\rangle$ corresponds to a free electron, while excited state $|e\rangle$ corresponds to the created trion. The total number of available excitations is thus equal to the number of free electrons $n_e \equiv N$. Using the analogy, we write the Hamiltonian of the system as
\begin{align}
\label{eq:H_gen}
\hat{H} = \omega_c \hat{a}^\dagger \hat{a} + \sum_j^{N} \Big\{ \Delta |e\rangle_j \langle e| + g(|e\rangle_j \langle g| \hat{a} + h.c.) \Big\},
\end{align}
where $\hat{a}^\dagger$ ($\hat{a}$) is a creation (annihilation) operator for a cavity mode, and $j$ corresponds to the considered two level system. The coupling term $\propto g$ thus describes polaritonic physics. For simplicity, we can take $\Delta = 0$ and measure cavity mode energy $\omega_c$ from this value.

The usual way to treat light-matter coupling in \eqref{eq:H_gen} is to assume weak excitation conditions and perform effective bosonization~\cite{PolzikRev} (i.e. make Holstein-Primakoff transformation). For this, the excitation creation operator reads
\begin{align}
\label{eq:b_dagger}
\hat{b}^\dagger = \frac{1}{\sqrt{N}} \sum_j |e\rangle_j \langle g|,
\end{align}
where $1/\sqrt{N}$ corresponds to the normalization condition, and $\hat{b}$ can be written similarly. The overall meaning of $\hat{b}^\dagger$ is the creation of excitation out available two-level emitters (free electrons) as a superposition. With the new operators Hamiltonian \eqref{eq:H_gen} can be recast in the familiar form
\begin{align}
\label{eq:H_pol}
\hat{H} = \omega_c \hat{a}^\dagger \hat{a} + g \sqrt{N} (\hat{b}^\dagger \hat{a} + h.c.),
\end{align}
where the last term corresponds to the usual polaritonic coupling with the superradiant enhancement, as for coupling to an ensemble of emitters. To see the influence of the light-matter coupling on the energy of the system (polaritonic shift), we take the many-body wave function in the form
\begin{align}
\label{eq:psi_G}
|\Psi_G\rangle = \{ |n_{ph}, n_{exc}\rangle, |n_{ph}, n_{exc} -1 \rangle, |n_{ph}-1, n_{exc}\rangle, |n_{ph}-1, n_{exc}-1\rangle, ... \},
\end{align}
where $n_{ph}$ corresponds to the number of photons and $n_{exc}$ to the number of excitations (i.e. number of $|1\rangle_j$ atomic states). Only certain states will be coupled by the off-diagonal light-matter interaction term. The expectation value for the Hamiltonian \eqref{eq:H_pol} yields
\begin{align}
\langle \hat{H} \rangle &= \omega_c n_{ph} + \langle n_{ph} -1, n_{exc} | g\sqrt{N} \hat{b}^\dagger \hat{a} |n_{ph},n_{exc}-1\rangle = \omega_c n_{ph} + g \sqrt{N} \langle n_{ph} -1, n_{exc} | \sqrt{n_{ph}} \sqrt{n_{exc}} |n_{ph}-1,n_{exc}\rangle \\ \notag &= n_{ph} (\omega_c + g\sqrt{N}),
\end{align}
where we considered number of excitations to be equal to number of photons. One observes that the expectation value contains the same value of coupling $g\sqrt{N}$ as before, which is equal to Rabi frequency.

The situation is however different when weak excitation conditions are not met. In this case effective bosonic picture (and Holstein-Primakoff transform) is not longer valid. To compare the two cases, the full basis in \eqref{eq:H_gen} must be considered. Choosing the wave function with $N-m$ states to excited, and $m$ states be in ground state $\{ 0_{i}^{m} \}$, we can write it as
\begin{align}
\label{eq:psi_E}
| \Psi_E \rangle = \{ |n_{ph} \rangle \otimes | 1_1, 1_2, .., 0_{i_1},.. 0_{i_2}, .., 1_N\rangle, ... \}.
\end{align}
We note that \eqref{eq:psi_E} is of course an approximation, and the full state may contain components with different number of excited stated. However, as the coherent state distribution is expected, their influence is suppressed.

Taking the expectation value for \eqref{eq:H_gen} for \eqref{eq:psi_E} one gets
\begin{align}
\label{eq:E_E}
\langle \hat{H} \rangle = \omega_c n_{ph} + \langle \Psi_E | \sum_j^{N} g|e\rangle_j \langle g| \hat{a} |\Psi_E \rangle = \omega_c n_{ph} + g \sqrt{n_{ph}} \sqrt{m} = n_{ph} (\omega_c + g\sqrt{m/n_{ph}}),
\end{align}
where the coupling reduced by the factor $\sqrt{m/N}$. For growing number of excitations $n_{exc}$ (and increasing number of atom in ground state), this prefactor can be expanded into series, leading to $\propto \sqrt{m/N} \approx 1 - 2 n_{exc} / N + O[n_{exc}/N]^2$ dependence. Finally, it is easy to see that for all states being excited ($m=0$) the coupling goes to zero, as there are no states to couple. This leads to the conclusion that for the inverted medium (maximal number of trions), the cavity becomes decoupled from the matter. Of course, in reality other effects coming from non-fermionicity play the role, and linear quech to zero is expected to change into a smooth function.

\subsection{Trion polaritons at increasing density} Finally, we describe the process of trion Rabi quench, evidenced by the behavior of the middle polariton branch. To describe Fig. 3(b) in the main text we perform the diagonalization of the full photon-trion-exciton system, which can be written as
\begin{align}
\label{eq:H3_matrix}
\mathbf{H}_{\mathrm{T-pol}} = \left( \begin{array}{ccc}
    E_{C} & \frac{1}{2}\Omega_{X}(n_X) & \frac{1}{2}\Omega_{T}(n_T) \\
    \frac{1}{2}\Omega_{X}(n_X) & E_X(n_X) & 0 \\
    \frac{1}{2}\Omega_{T}(n_T) & 0 & E_T \end{array} \right).
\end{align}
Here we consider the nonlinear contributions to both trion and exciton modes, taking experimentally measured values $E_T^{(0)} = 1621.02$~meV, $E_C = 1630.32$~meV, $E_X^{(0)} = 1645.72$~ meV, $\Omega_T = 5.8$~meV, $\Omega_X=17.2$~meV, obtained exciton Bohr radius $a_B = 0.85$~nm, variation parameters $\lambda_1 \approx \lambda = 0.87$~nm, $\lambda_2 \approx \lambda'=2.54$~nm, and using the estimated electron concentration $n_e = 4.05 \times 10^{10}$~cm$^{-2}$. The result is shown in Fig.~3b [main text] by the red solid curve, and allows to reproduce the initial red shift of $E_{\mathrm{MPB}}$  due to quenched trion Rabi-splitting, followed by the weak blue shift caused by exciton-exciton interactions. 

Finally, we notice that in experiment the energy blue shift of the MPB at small exciton concentrations (see Figs.~3b and 5 of the main text) is about one order of magnitude larger than that predicted by the theory. The missing blueshift contribution, which manifests as a plateau due to competition with $\Omega_{T}$ quench and subsequent linear growth, is identified as a trion-exciton interaction. Indeed, the full trion Rabi quench appears when high order terms are neglected, with function $f_T(n_T)$ reaching zero non-smoothly. However, their account shall provide non-zero $\Omega_{T}$ even for $n_T \approx n_e/2$, leading to residual coupling and small trion admixture. This for instance can lead to several percent fraction of the trion in MPB branch, and contribute as trion-exciton exchange.  The interaction can substantially increase the energy of MPB state even for the trion being weakly coupled to light. As a result, at exciton densities 5-10 times  above the maximum density of the excited trions, the MPB can exhibit substantial blueshift well above that predicted by the theory, which considers only neutral exciton-exciton interactions. 

The full calculation of trion-exciton interaction energy is formidable and is typically limited due to its large dependence on the trion wavefunction ansatz \cite{Combescot2011}. Nevertheless, the overall strength of trion-exciton interaction is estimated to be one-to-two orders of magnitude stronger than that of exciton-exciton exchange. This is because of the following reasons: 1) Firstly, the number of electron and hole exchange processes is increased compared to the X-X exchange case; 2) the outer shell of the trion, described by $\lambda_2$, defines the  scattering cross-section, which is larger than for exciton; 3) the direct Coulomb term is expected to play a role, unlike for the neutral exciton-exciton interaction case.

{\color{black}

\section{Quantum trion-polaritons: theory}
\label{sec:QT}

We now present a theoretical estimate of how a system analogous to the one considered in the current paper can be used to observe strong nonlinear response at the quantum level of a few photons. The experimental signature of such behavior is a pronounced antibunching of the photon emission, which can be monitored by measuring the second order coherence function. To calculate the latter, we consider a system described by the trion-photon Hamiltonian, where the coupling is between a single photonic mode of the cavity, described by the bosonic operators $\hat{c},\hat{c}^\dagger$, and a trion mode with zero in-plane momentum $\textbf{K}=0$, characterized by the operators $\hat{B}_0,\hat{B}_0^\dagger$ [see definition in Eq.~\eqref{eq:B_K}]. The Hamiltonian can be written in the rotating frame as 
\begin{align}
    \label{eq:Hamiltonian_quantum}
    \hat{H} = \omega_c \hat{c}^\dagger \hat{c} + \omega_T \hat{B_0}^\dagger
    \hat{B_0} + \frac{\Omega_T}{2}(\hat{B}_0^\dagger \hat{c} + \hat{B}_0 \hat{c}^\dagger) + P (\hat{c}e^{i\omega_p t} + \hat{c}^\dagger e^{-i\omega_p t}),
\end{align}
where $P$ denotes the pump strength for a cw coherent optical drive of frequency $\omega_p$. In Eq.~\eqref{eq:Hamiltonian_quantum} the first two terms correspond to free cavity photons and trions and the third term describes the photon-trion coupling. The Rabi splitting $\Omega_T$ is given by Eq.~\eqref{eq:trion_Rabi}. The effective polariton energies in the weak excitation limit read $\omega_{L,U} = (\omega_c + \omega_T)/2 \mp \sqrt{\Omega_T^2 + (\omega_c - \omega_T)^2}/2$. Finally, it is convenient to go to the rotating frame with respect to the last term in Eq. \eqref{eq:Hamiltonian_quantum}, such that the system is described by the detuning from the pump frequency.

The non-bosonic operator $\hat{B}_0^\dagger$ is characterized by its matrix elements in the Hilbert space spanned by the Fock states for trions, $|N_T\rangle$. The matrix elements read
\begin{align}
     \langle N_T-1| \hat{B}_{0} |N_T\rangle =  \frac{\langle \text{\O} |\hat{B}_{0}^{N_T-1} \hat{B}_{\mathbf{q}\rightarrow 0} (\hat{B}_{0}^\dagger)^{N_T}| \text{\O}\rangle}{\sqrt{(N_T-1)! F_{N_T-1}} \sqrt{N_T! F_{N_T}}},
    \label{eq:off-diag_split}
\end{align}
where $F_{N_T} \equiv \langle \text{\O} |\hat{B}_{0}^{N_T}  \hat{B}_{0}^{\dagger N_T}| \text{\O}\rangle/N_T!$ corresponds to the correction factor accounting for the composite nature of the trion excitation (in the case of bosonic excitations $F_{N_T}=1$).  Its explicit form can be derived using the recursive relation
\begin{align}
    \label{eq:F_N_rec}
    F_{N_T - n}= \frac{(N_e - N_T)! N_e^{n}}{(N_e-N_T+n)!} F_{N_T},
\end{align}
which is similar to those reported earlier for Frenkel excitons \cite{Pogosov2009} and intersubband excitations \cite{DeLiberato2009}.
\begin{figure}
\centering
\includegraphics[width=0.9\linewidth]{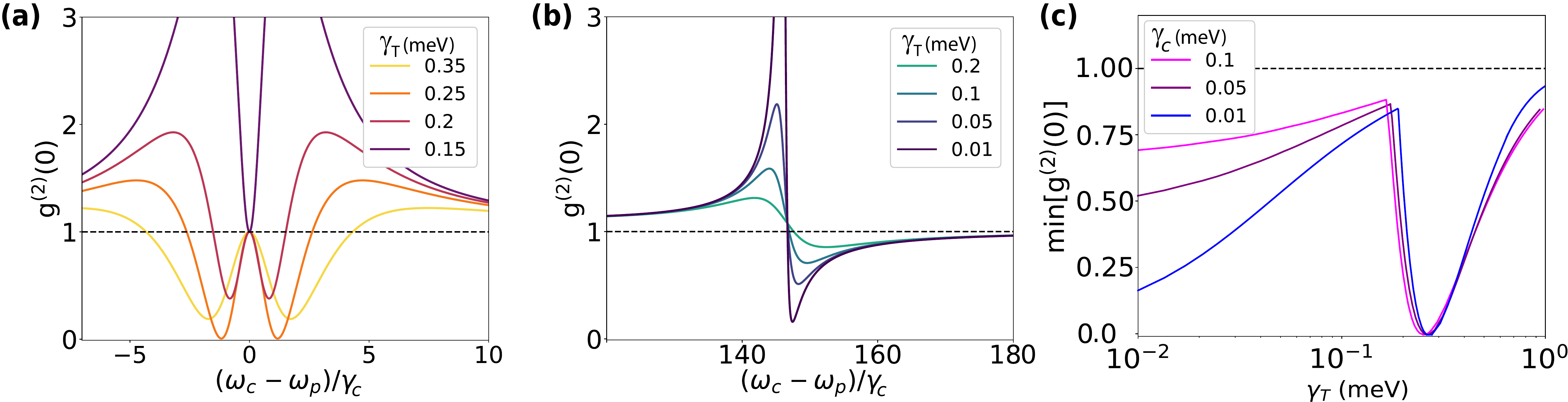}
\caption{{\color{black} Second order coherence at zero delay for the MoSe$_2$ trion-polariton system. \textbf{(a)} $g^{(2)}(0)$ as a function of pump detuning plotted in the vicinity of the cavity resonance (we consider $\omega_c = \omega_T$), for several values of the nonradiative decay rate $\gamma_T$ of the trion mode ($N_e = 100$). The destructive interference leads to the appearance of antibunching as $\gamma_T \sim g_c := \Omega_T/\sqrt{N_e}$, corresponding to unconventional photon blockade. \textbf{(b)} $g^{(2)}(0)$ for pump frequencies close to the lower polariton frequency, $\omega_p \sim \omega_L$. Conventional photon blockade leads to the appearance of antibunching at small $\gamma_{T,c}$. \textbf{(c)} The minimal $g^{(2)}(0)$, minimized over pump detuning, and shown as a function of trion decay $\gamma_T$ for several cavity quality factors. Qualitatively different behaviour is visible for $\gamma_{T} \ll g_c = 0.29$~meV and $\gamma_{T} \sim g_c$.} }
\label{fig:g2}
\end{figure}
As for the numerator, it can be evaluated using the commutation relation
\begin{align}
    \label{eq:Bq_Bq_dag_N}
    [\hat{B}_{\mathbf{q}}, \hat{B}_{\mathbf{q'}}^{\dagger N}] = N (\hat{B}_{\mathbf{q}'}^\dagger)^{N-1} (\delta_{\mathbf{q},\mathbf{q}'} - \hat{D}_{\mathbf{q},\mathbf{q}'}) - \frac{N (N-1)}{N_e} \hat{B}_{2\mathbf{q}' - \mathbf{q}}^\dagger (\hat{B}_{\mathbf{q}'}^\dagger)^{N-2},
\end{align}
derived iteratively from Eq.~\eqref{eq:comm_B}. A careful treatment of recursion in arbitrary order gives a closed expression for the matrix elements 
\begin{align}
\label{eq:off-diag}
    \langle N_T - 1| \hat{B}_{0} | N_T \rangle = \sqrt{N_T \left( 1 - \frac{N_T}{N_e + 1}\right)} \sqrt{\frac{N_e}{N_e + 1}} \left[ 1 - (-1)^{N_T} \frac{(N_e - N_T)! N_T!}{N_e!}\right],
\end{align}
and a similar expression can be derived for its complex conjugate. Importantly, Eq.~\eqref{eq:off-diag} works for the relevant case of few trion excitations $N_T \leq N_e$ (contrary to the Holstein-Primakoff approach \cite{Holstein,Emary2003,Brandes2003} which fails in the limit where $N_T = 1$). This will be important for obtaining correct quantum statistical properties for the cavity emission.

The dynamics is studied by numerically solving the master equation for the full density matrix of the system in the truncated trion-photon Hilbert space. It reads
\begin{equation}
\frac{\partial\hat{\rho}}{\partial t}=i[\hat{\rho},\hat{H}]+\gamma_{c}\left[\hat{c}\hat{\rho}\hat{c}^\dagger-\frac{1}{2}\left(\hat{c}^\dagger \hat{c}\hat{\rho}+\hat{\rho}\hat{c}^\dagger \hat{c}\right)\right]+\gamma_{T}\left[\hat{B}_0\hat{\rho}\hat{B}_0^\dagger-\frac{1}{2}\left(\hat{B}_0^\dagger \hat{B}_0\hat{\rho}+\hat{\rho}\hat{B}_0^\dagger \hat{B}_0\right)\right],
\end{equation}
where the first term on the right hand side corresponds to the coherent part of the evolution, the second term describes photonic dissipation (characterized by the finite broadening of the cavity mode related to the finite lifetime of the cavity photons), $\gamma_{c}=\tau_{c}^{-1}$, and the third term describes trion dissipation characterized by nonradiative broadening $\gamma_{T}=\tau_{T}^{-1}$.

To characterize the statistics of the cavity output we evaluated the second order coherence function at zero delay for the cavity photons, defined as
\begin{equation}
g^{(2)}(0) = \frac{\text{Tr}\left[\hat{c}^\dagger \hat{c}^\dagger \hat{c} \hat{c}\hat{\rho}_s\right]}{\text{Tr}\left[\hat{c}^\dagger \hat{c}\hat{\rho}_s\right]^2},
\end{equation}
where $\hat{\rho}_s$ denotes the steady state density matrix for the continuously driven dissipative system. 

In the current experiment we measured a Rabi frequency $\Omega_T = 5.8$~meV and estimated the free electron density $n_e = 4\times 10^{10}$~cm$^{-2}$, with the corresponding number of electrons being $\sim 1200$ in the cavity of area $A = 3$ $\mu$m$^2$. These experimental values already imply the expected trion-photon coupling strength for the case of a single electron in the cavity area, $g_c = \Omega_T/\sqrt{N_e} = 0.17$~meV. In improved devices the photonic mode area can potentially be reduced to $A = 1~\mu$m$^2$ using a curved top mirror with smaller radius of curvature \cite{Dufferwiel2014,Dufferwiel2015c} so that the coupling strength is enhanced by a factor of $\sqrt{3}$, giving $g_c = 0.29$~meV. Meanwhile, the photon decay rates for an open cavity routinely reach $\sim 0.1$~meV values, and can be as low as $\gamma_c = 10~\mu$eV ($\sim 65$ ps lifetime) in the state-of-the-art samples \cite{Besga2015,Dufferwiel2015c}. 
Furthermore, electron concentrations down to $n_e = 10^{10}$ cm$^{-2}$ can be realised in gated samples. In this case the expected Rabi splitting will be $\sim 3$~meV for an average cavity occupation of 100 electrons, making phase-space filling effects even more pronounced and going far beyond the electronic confinement regime. Finally, in TMDC samples of high purity, which are encapsulated between thick hBN-layers, the trion inhomogeneous broadening as well as non-radiative recombination may become negligible such that the trion nonradiative linewidth $\gamma_T$ will be determined by pure dephasing due to scattering with phonons\cite{Martin2018}. This may result in nonradiative trion linewidths as small as $\gamma_T\sim10~\mu$eV at a temperature $T$ of 1 K \cite{Martin2018}.


{\color{black}While studying the second-order coherence in the system, we note two possible mechanisms which can reduce the multi-photon component and facilitate single photon emission. The first mechanism corresponds to the conventional blockade-type antibunching, where two-photon occupation is suppressed by strong trion-photon coupling at the single particle level, $g_c/\gamma_{c,T} \gg 1$. In the following we show that this shall be possible in future high-quality samples. The second mechanism can be identified as an unconventional-type single photon blockade \cite{Liew2010,Bamba2011} due to phase space filling effects, which does not require strong coupling and works at optimal parameters of $g_c \sim \gamma_T$ and $\omega_p \approx \omega_c$ (see \cite{Kyriienko2019} for the full analysis). It relies on destructive interference between the direct coherent optical excitation two photons and the trion-mediated excitation path \cite{Bamba2011}, thus relaxing the requirement for strong energy shift. At the same time, it causes oscillations of the second-order coherence as a function of delay, and generally has smaller emission probability. Below, we consider the two regimes as long-term and near-term goals for nonlinear trion-polaritonics with TMDC materials.

The results of the second-order coherence calculations are shown in Fig.~\ref{fig:g2}. First, in Fig.~\ref{fig:g2}(a) we plot $g^{(2)}(0)$ for a range of pump frequencies close to the cavity transition. For this, we consider cavity linewidth $\gamma_c = 0.05$~meV, $\omega_c = \omega_T$, and $g_c = 0.29$~meV with 100 electrons. Studying the dependence for different values of the nonradiative trion decay rate $\gamma_T$ we observe the appearance of an antibunching window when $g_c \approx \gamma_T$. At the same time, for narrow linewidths $\gamma_T \ll g_c$ the antibunching behaviour disappears from the $\omega_p \approx \omega_c$ region, signifying the resonant interference-based nature of the effect and the modest coupling requirement. At the same time, we note the limited efficiency of the single photon emission in this window, as cavity occupation is typically in $\langle \hat{c}^\dagger \hat{c} \rangle \sim 10^{-3}..10^{-4}$. We envisage that in practise the optimal parameters would be tuned using sample positioning in an open cavity as a tool.

In Fig.~\ref{fig:g2}(b) we consider a different range of pump detunings, where the coherent drive is nearly resonant with the lower trion-polariton, $\omega_p \approx \omega_L$, which corresponds to non-zero $\omega_c - \omega_p$ detuning. In the calculations we assume a high-quality cavity with $\gamma_c = 10~\mu$eV and improved values of the trion linewidth limited by thermal effects. The plot for $g^{(2)}(0)$ shows an antibunching for pump frequencies slightly below the transition, with single photon purity gradually improving as the coupling ratio $g_c/\gamma_{c,T}$ increases. The Fano-shape profile of the $g^{(2)}(0)$ dependence draws the connection to conventional Kerr-based polariton blockade \cite{Verger2006}, which can be accessed in structures of larger lateral size and favours the strong binding energy limit for excitons.

Finally, Fig.~\ref{fig:g2}(c) shows the minimal value of $g^{(2)}(0)$, minimized over a wide range of pump detunings spanning both regimes, as a function of the trion nonradiative linewidth $\gamma_T$. For trion decay rate comparable to the light-matter coupling constant $g_c$ the unconventional antibunching can be observed ($\gamma_T > 0.1$~meV). For long-lived trions with small non-radiative decay (0.01 meV) we also observe pronounced antibunching due to conventional blockade, which is limited by the cavity quality factor. This shows that single photon emission with trion-polaritons in MoSe$_2$ is possible in high quality samples.
}


}
